\documentclass{aa}

\usepackage{graphicx}
\usepackage{txfonts}
\usepackage[colorlinks,linktocpage=true]{hyperref}

\begin{document} 

\title{Probing modified Newtonian dynamics \\ with hypervelocity stars}

   \author{Sankha Subhra Chakrabarty\thanks{sankhasubhra.chakrabarty@unito.it}\inst{1,2}
          \and
          Luisa Ostorero\inst{1,2}
                  \and
                  Arianna Gallo\inst{1,2}
                  \and
                  Stefano Ebagezio\inst{3,1}
                  \and
                  Antonaldo Diaferio\inst{1,2}}

   \institute{Dipartimento di Fisica, Universit\`a di Torino,  Via P. Giuria 1, I-10125 Torino, Italy
         \and
             Istituto Nazionale di Fisica Nucleare (INFN), Sezione di Torino, Via P. Giuria 1, I-10125 Torino, Italy 
        \and I.~Physikalisches Institut, Universit\"at zu K\"oln , Z\"ulpicher Str.\ 77, D-50937 K\"oln, Germany}

   \date{Received date; accepted date}
 
  \abstract
  {
  We show that measuring the velocity components of  hypervelocity stars (HVSs) can discriminate between modified Newtonian dynamics (MOND) and Newtonian gravity. Hypervelocity stars are ejected from the Galactic center on radial trajectories with a null tangential velocity component in the reference frame of the Galaxy. They acquire tangential components due to the nonspherical components of the Galactic gravitational potential. Axisymmetric potentials only affect the latitudinal components, $v_\theta$, and non-null azimuthal components, $v_\phi$, originate from non-axisymmetric matter distributions. For HVSs with  sufficiently high ejection speed, the azimuthal velocity components are proportionate to the deviation of the gravitational potential from axial symmetry. The ejection velocity threshold is $\sim$~750~km~s$^{-1}$ for 4~$M_{\sun}$ stars and increases with decreasing HVS mass. We determine the upper limit of $v_\phi$ as a function of the galactocentric distance for these high-speed HVSs if MOND, in its quasi-linear formulation QUMOND, is the correct theory of gravity and either the triaxial Galactic bulge or a nonspherical hot gaseous halo is the primary source of the azimuthal component, $v_\phi$. In Newtonian gravity, the HVSs within 60 kpc of the Galactic center may easily have $v_\phi$ values higher than the QUMOND upper limit if the dark matter halo is triaxial or if the dark matter halo and the baryonic components are axisymmetric but their two axes of symmetry are misaligned. Therefore, even a limited sample of high-speed HVSs could in principle allow us to distinguish between the QUMOND scenario and the dark matter model. This test is currently limited by (i) the lack of a proper procedure to assess whether a star originates from the Galactic center and thus is indeed an HVS in the model one wishes to constrain; and (ii)  the large uncertainties on the galactocentric azimuthal velocity components, which should be reduced by at least a factor of $\sim 10$ to make this test conclusive. A proper procedure to assess the HVS nature of the observed stars and astrometric measurements with microarcsecond precision would make  this test feasible.
  }

   \keywords{Gravitations --
            Cosmology: dark matter --
            Galaxy: general --
                        Galaxy: structure --
                        Galaxy: kinematics and dynamics
            }

   \maketitle


\section{Introduction}
\label{sec:intro}

The mass discrepancy in the Universe originates from a set of independent observations on galactic and cosmological scales. On the scale of galaxies, nearly flat rotation curves of disk galaxies at large radii \citep{1978PhDT.......195B, 1982ApJ...261..439R, 1985ApJ...289...81R, 1996MNRAS.281...27P, McGaugh:2000sr, 2001ARA&A..39..137S, 2013A&A...557A.131M, 2014ApJ...785...63B}, the stability of dynamically cold stellar disks \citep{1971ApJ...168..343H, 1973ApJ...186..467O, 1980MNRAS.193..189F, 1991Natur.352..411R, 2014RvMP...86....1S, 2019MNRAS.486.4710S}, the dynamics of the outer regions of ellipticals \citep{deZeeuw1991, 1991ApJ...383..112F, 2006MNRAS.366.1126C, 2013MNRAS.428..389P, 2016ARA&A..54..597C, 2018A&A...618A..94P}, and the large mass-to-light ratios of dwarf galaxies \citep{kormendy1987, 1995MNRAS.277.1354I, 1998ARA&A..36..435M, 2001MNRAS.325.1017V, 2009ApJ...704.1274W, 2009MNRAS.394L.102L, 2019ARA&A..57..375S} cannot be explained in the standard theory of gravity without assuming the existence of a large amount of dark matter \citep{1984Natur.311..517B, 1991ApJ...379...52W, 2018ARA&A..56..435W, 2019Galax...7...81Z}. On the scale of galaxy clusters \citep{2000cucg.confE...1B, 2005RvMP...77..207V, 2008SSRv..134....7D, 2019SSRv..215....7W}, the dynamics of the member galaxies \citep{1996ApJ...471..643K, 2009ApJ...703..982G, 2013ApJ...764...58G, 2013ApJ...767...15R, 2017ApJS..229...20S, 2020arXiv201000992T}, the X-ray emission of the intracluster gas \citep{1986RvMP...58....1S, 2002ARA&A..40..539R, 2010A&ARv..18..127B, 2019ApJ...880..142S, 2020MNRAS.497.3976C}, and gravitational lensing \citep{1990ApJ...349L...1T, 2012ApJS..199...25P, 2018NatAs...2..744S, 2020Sci...369.1347M, 2020A&ARv..28....7U} imply mass-to-light ratios of order $\sim 100-400~M_\sun/L_\sun$  \citep{2002ApJ...569..720G, 2004AJ....128.1078R, 2015MNRAS.449.2345P}. On  cosmological scales, the structure formation from the nearly homogeneous matter distribution implied by the small temperature anisotropies of the cosmic microwave background (CMB) requires a stronger gravitational pull than provided by the baryonic matter alone \citep{1967Natur.215.1155S, 1985ApJ...292..371D, 2007ApJS..170..377S, 2020A&A...641A...6P}. 

The most popular and widely investigated solution to the problem of the mass discrepancy is to assume the presence of cold dark matter (CDM) that is non-baryonic and interacts with the baryons only via gravity \citep{1982ApJ...263L...1P, 1982ApJ...255..341B, 1982Natur.299...37B, 1984Natur.311..517B, 2007ApJS..170..377S, 2012AnP...524..507F, 2013PhR...531....1S, 2020A&A...641A...6P}. However, to date, none of the elementary particles suggested as candidates of dark matter has been detected \citep{2018PhRvD..98c0001T}. The allowed windows for the parameters associated with various dark matter particles are also shrinking due to constraints from various terrestrial experiments and from astrophysical observations \citep{Bertone:review, deMartino:review}. 

In principle, the mass discrepancy problem can be solved with a modification of the theory of gravity rather than with dark matter \citep{1990A&ARv...2....1S, 2002ARA&A..40..263S, 2012PhR...513....1C, 2017PhR...692....1N, deMartino:review}. Modified Newtonian dynamics \citep[MOND;][]{MOND:Milgrom, 1983ApJ...270..371M, 1983ApJ...270..384M} is one of the most investigated modifications of Newtonian gravity \citep{Famaey:nux, 2020Galax...8...35M, deMartino:review}. Modified Newtonian dynamics suggests that Newtonian gravity breaks down in the low acceleration regime, where the gravitational field is smaller than the critical value $a_0\approx 10^{-10}$~m~s$^{-2}$. Due to its dependence on the acceleration, the formulation of a covariant version of MOND, which is required to test the theory against the  properties of the large-scale structure on cosmic scales,  is not unique and remains difficult 
\citep{1984ApJ...286....7B, 2004PhRvD..70h3509B, 2009PhRvD..80l3536M,  2015CaJPh..93..107M, hernandez2019, 2020arXiv200700082S}.

On the other hand, on the scale of galaxies, MOND has proved to be predictive and successful \citep[e.g.,][]{merritt2020book}: Confirmed MOND predictions  include the baryonic Tully-Fisher relation of disk galaxies with virtually no systematic scatter
over $\sim 5$ orders of magnitude in baryonic mass \citep{McGaugh:2000sr, 2016ApJ...816L..14L, 2020Galax...8...35M}, the dependence of the shape of the rotation curves on the surface brightness of the galaxy \citep{2020Galax...8...35M}, and the large mass-to-light ratios of dwarf galaxies when interpreted in Newtonian gravity \citep{aaronson1983,kormendy1987, 1991AJ....102..914M, 1998ARA&A..36..435M}; as expected in MOND, the strong equivalence principle also appears to be invalid, as suggested by the analysis of accurate rotation curves of disk galaxies   \citep{2020ApJ...904...51C}.

A number of MOND predictions concerning the Milky Way \citep[MW;][]{Famaey:nux} and the nearby dwarfs \citep{2020A&A...640A..26H} still need to be tested with upcoming astrometric data. Indeed, when interpreted in Newtonian gravity, the MOND gravitational field requires the presence of ``phantom dark matter'' in addition to the baryonic matter
\citep{1983ApJ...270..371M, 1984ApJ...286....7B}. Specifically,  MOND predicts the existence of a disk of phantom dark matter in the MW \citep{2009AA...500..801B}; therefore, the acceleration of the stars perpendicular to the plane of the disk can constrain the total surface density of the baryonic and phantom disks \citep{2007MNRAS.379..597N}. At 1.1 kpc above the Galactic plane at the distance of the Sun from the Galactic center, the total surface density in MOND is expected to be 60\% larger than the surface density of the baryonic disk, whereas this enhancement is predicted to be 51\% in Newtonian gravity with a spherical dark matter halo \citep{2009AA...500..801B, Famaey:nux}. Similarly, the scale length of the ``baryonic + phantom'' disk can be constrained by the vertical acceleration profiles at different radii. In the formulation of MOND suggested by \cite{1984ApJ...286....7B}, this scale length should be 1.25 times the scale length of the visible stellar disk \citep{2009AA...500..801B, Famaey:nux}. Moreover, the angle between the minor axis of the velocity ellipsoid of the stars in the solar neighborhood and the vertical direction depends on the shape of the gravitational potential \citep{1991MNRAS.253..427C}: For axisymmetric potentials, the angles expected in MOND and Newtonian gravity with a spherical dark matter halo differ by 2 degrees at the distance of the Sun from the MW center and at 2 kpc above the MW disk \citep{2009AA...500..801B}.

In this work we propose a novel prediction of MOND  concerning the positions and proper motions of  hypervelocity stars (HVSs) within the MW. The existence of HVSs was predicted by Hills (\citeyear{Hills88}) as a result of three-body interactions between the supermassive black hole (SMBH) at the Galactic center and binary stars. After the interactions, one of the binary stars is ejected with high radial velocity while the other is captured by the SMBH. Some other mechanisms of generating HVSs have been proposed, including the interactions between a star and the binary black hole at the center \citep{YuTremaine}, the interactions between binary stars and binary black holes \citep{2018MNRAS.475.4595W}, the interactions between the MW and a dwarf galaxy \citep{Abadi}, and star formation in the outflows driven by an active galactic nucleus \citep{WangLoeb, Silketal}. Since the serendipitous discovery of the first HVS \citep{brown2005}, $\sim 90$ high-velocity stars have been identified as candidate HVSs  \citep[see, e.g.,][]{hirsch2005, edelmann2005, brown2006, brown2009, brown2012, brown2014, brown2018, 2011A&A...527A.137T, 2012ApJ...744L..24L, 2015RAA....15.1364L, 2013A&A...559A..12P, 2014ApJ...785L..23Z, 2017ApJ...847L...9H, 2017MNRAS.470.1388M, 2018AJ....155..207N, 2018AJ....156..265M,  2018ApJ...866..121H, 2019MNRAS.483.2007E, 2019ApJS..244....4D, 2019ApJ...887L..39L, koposov2020, Li2021}.

The HVSs are ejected from the Galactic center with a speed of  several hundred km~s$^{-1}$ or higher. These stars are sometimes separated into samples of bound and unbound stars. This distinction is irrelevant for our purpose here, and we consider as an HVS any star that is ejected from the Galactic center on a purely radial orbit with null tangential velocity in the galactocentric reference frame. Hypervelocity stars obtain nonzero tangential velocities due to the nonspherical components of the Galaxy gravitational potential. As they travel to large distances from the center, the distribution of their positions and velocities carries signatures of the gravitational field of the Galaxy. Several schemes to constrain the shape of the halo \citep{2005ApJ...634..344G, 2017MNRAS.467.1844R, Contigiani:2018axw} and to measure the virial mass of the MW \citep{Fragione_2017} using the HVSs have been proposed. \cite{2009ApJ...697.2096P} suggested a novel method to discriminate between various models of Galactic potential within CDM and MOND paradigms by using the asymmetry of the velocity distributions of incoming and outgoing HVSs. \cite{2018ApJ...869...33H} also proposed a method for estimating the position and velocity of the Sun within the MW by using the fact that the HVSs have low tangential velocities. \cite{kenyon2018} showed that tangential velocities may also be caused by systems external to the MW, such as the Large Magellanic Cloud (LMC).

In this work we simulate the kinematics of the HVSs in MOND as well as in Newtonian gravity. We show that the azimuthal components, $v_\phi$, of the tangential velocities  of the HVSs may distinguish MOND from Newtonian gravity. Section \ref{sec:modg} illustrates the quasi-linear formulation of MOND (QUMOND) that we adopt in this work. Section \ref{sec:pot} describes our model of the distribution of the MW baryonic matter that generates the QUMOND gravitational potential.  In Sect.~\ref{sec:dark} we illustrate the model of the dark matter halo we adopt for comparison with the QUMOND predictions. In Sect.~\ref{sec:HVS} we illustrate and discuss our simulations. In Sect.~\ref{sec:angvel} we show the galactocentric tangential velocities of the HVSs in QUMOND and detail our QUMOND predictions. We conclude in Sect.~\ref{sec:conclusions}.

\section{Quasi-linear modified Newtonian dynamics}
\label{sec:modg}

Throughout this work, we adopted QUMOND, the quasi-linear formulation of MOND \citep{QUMOND:Milgrom}, where the gravitational field is 
\begin{equation}
    \vec{g} = \nu \left( \frac{|\vec{g}_{\rm N}|}{a_0} \right) \ \vec{g}_{\rm N} \,  \label{eq:QuMOND}
\end{equation}
and $\vec{g}_{\rm N}$ is the Newtonian gravitational field due to the baryonic matter alone. The interpolating function $\nu (x)$ satisfies the limits $\nu (x) \to 1$ when $x \gg 1$, and $\nu (x) \to x^{-1/2}$ when $x \ll 1$. We adopt
\begin{equation}
    \nu (x) = \left[ \frac{1}{2} \left( 1 + \sqrt{1 + 4 x^{-\gamma}} \right) \right]^{\frac{1}{\gamma}}\, ,  \label{eq:nu}
\end{equation}
with $\gamma$ = 1 or  $\gamma = $ 2 \citep{Famaey:nux}. The function with $\gamma$ = 1 is known as the simple interpolation function. The acceleration scale, below which Newtonian gravity  modifies, is set by $a_0$. The value of $a_0$ is found by fitting the rotation curve data \citep{1991MNRAS.249..523B, 2002A&A...393..453B}, the observed correlation between the mass discrepancy and the acceleration \citep{2004ApJ...609..652M}, and the baryonic Tully-Fisher relation \citep{McGaugh:galaxies}. The best-fit value of $a_0$ varies from 3000 to 4000 km$^2$ s$^{-2}$ kpc$^{-1}$ and also slightly depends on the chosen interpolation function $\nu(x)$ (Eq.~\ref{eq:nu}). We chose an intermediate value: $a_0 = 3600$ km$^2$ s$^{-2}$ kpc$^{-1}= 1.2 \times 10^{-10}$ m s$^{-2}$. 

For the Galaxy, we first assumed a simple model with three baryonic components: a central SMBH, a stellar disk and a stellar bulge. With this model, the Newtonian acceleration entering Eq.~(\ref{eq:QuMOND}) is 
\begin{equation}
        \vec{g}_{\rm N} = - \vec{\nabla} (\Phi_{\rm BH} + \Phi_{\rm Bulge} + \Phi_{\rm Disk}) \, , \label{eq:gN}
\end{equation}
where $\Phi_{\rm BH}$, $\Phi_{\rm Bulge}$, and $\Phi_{\rm Disk}$ are the Newtonian gravitational potentials that we provide in the next section. A more sophisticated model that includes the additional baryonic component of a hot gaseous (HG) halo will be discussed in Sect.~\ref{sec:HG}.

\section{Newtonian gravitational potentials}
\label{sec:pot}

We investigated two variants of the model for the baryonic gravitational potential that enters Eq.~(\ref{eq:gN}): an axisymmetric model and a non-axisymmetric model. In the latter model, the deviation from the axial symmetry originates only from the presence of a triaxial bulge. We used the reference frame of the Galaxy with the origin at the Galaxy center. We used cylindrical coordinates $(R, \phi, z)$ for the axisymmetric model and Cartesian coordinates $(x, y, z)$ for the triaxial model; $R$ lies on the $x$-$y$ plane, which we take as the equatorial plane of the Galactic disk. For the spherically symmetric components, we used spherical polar coordinates $(r, \theta , \phi)$.

\subsection{The axisymmetric model}
\label{sec:axisymm}

We adopted simple analytical potentials for the three MW components. For the stellar disk, we used
the Miyamoto-Nagai model \citep{Dsk:Miyamoto}
\begin{equation}
\Phi_{\rm Disk} (R,z)= - \frac{G M_{\rm D}}{\sqrt{R^2 + \left( a_{\rm D} + \sqrt{z^2 + b_{\rm D} ^2} \right)^2}} \label{eq:pot_disk}\, ,
\end{equation}
with $M_{\rm D} = 1.0 \times 10^{11} \ M_\sun $, $a_{\rm D} = 6.5 \ {\rm kpc}$, and $b_{\rm D} = 0.26 \ {\rm kpc}$ \citep{2014ApJ...794...59K, MW:Price-Whelan, 2017MNRAS.467.1844R, Contigiani:2018axw}. 

For the bulge, we took the  Hernquist sphere \citep{Blg:Hernquist}
\begin{equation}
\Phi_{\rm Bulge}(r) = - \frac{G M_{\rm B}}{r_{\rm B} + r} \label{eq:pot_bulge}\, ,
\end{equation}
with $M_{\rm B} = 3.4 \times 10^{10} \ M_\sun$, $r_{\rm B} = 0.7 \ {\rm kpc}$ \citep{2014ApJ...794...59K, MW:Price-Whelan, 2017MNRAS.467.1844R, Contigiani:2018axw}, and a total bulge mass of $3.0 \times 10^{10} M_\sun$. 

Finally, the gravitational potential of the central SMBH is \begin{equation}
\Phi_{\rm BH} (r) = - \frac{G M_{\rm BH}}{r} \label{eq:pot_BH}\, ,
\end{equation}
with $M_{\rm BH} = 4.0 \times 10^{6} \ M_\sun$. This mass of the SMBH is comparable to various estimates reported in the literature \citep[e.g.,][]{BH:Eisenhauer, BH:Ghez, 2016ApJ...830...17B, 2017ApJ...837...30G}. 

In this model, the axial symmetry of the MW originates only from the stellar disk because the gravitational potentials of the SMBH and the bulge are spherically symmetric.

\begin{figure}[!ht]
\centering 
\includegraphics[width=8cm]{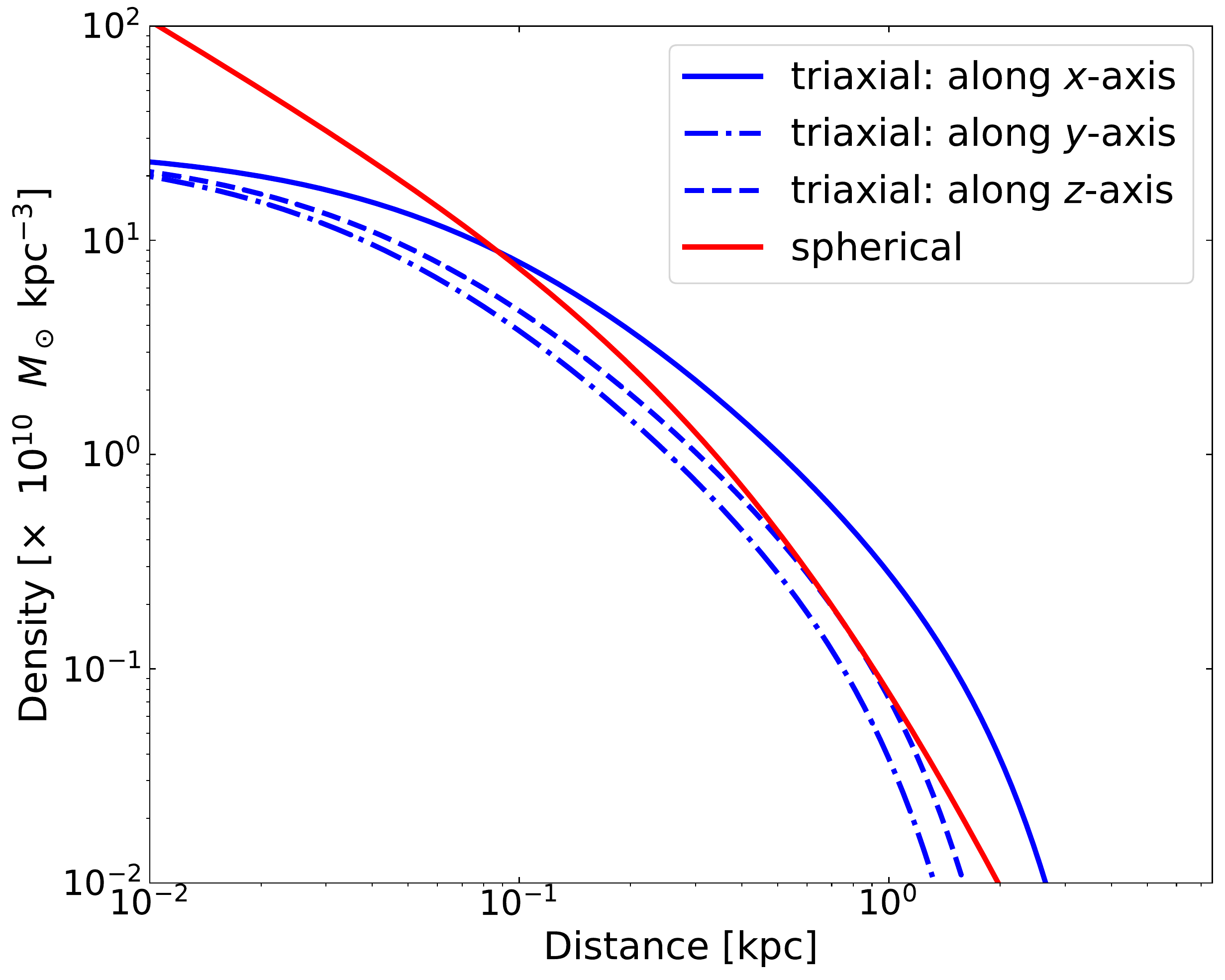}
\caption{\label{fig:bulgedens}  Density profiles of the spherical (red) and triaxial (blue) bulges as functions of the distance from the center. For the triaxial bulge, we show the density profiles along the $x$, $y$, and $z$ axes.}
\end{figure}

\begin{figure}[!ht]
\centering 
\includegraphics[width=8cm]{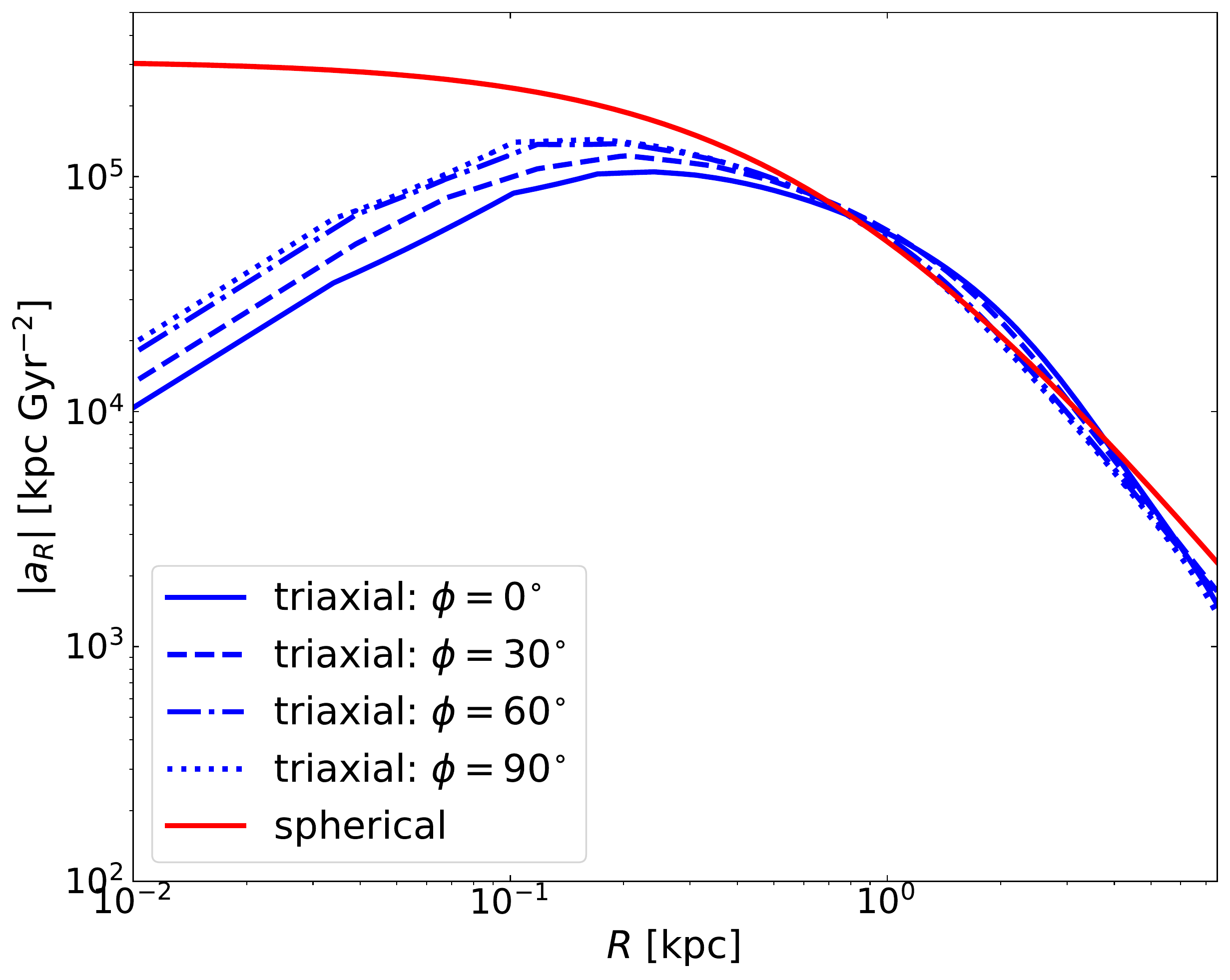}
\caption{\label{fig:bulgeacc}  Magnitude of the radial acceleration in the plane of the disk, $\vert a_R \vert$, due to the bulge alone in Newtonian gravity: spherical bulge (red) and triaxial bulge (blue). For the triaxial bulge, the acceleration varies with the azimuthal angle, $\phi$, which is the angle with respect to the $x$ axis in the plane of the disk. We show the results for various values of $\phi$; $\phi = 0^\circ$ and $\phi=90^\circ$ correspond to the $x$ and $y$ axes, respectively.}
\end{figure}

\subsection{The non-axisymmetric model: The triaxial bulge}
\label{sec:bulge}

To estimate the impact of a non-axisymmetric distribution of the baryonic matter on the HVSs kinematics in QUMOND, we considered the effects of a triaxial bulge and kept unaltered the gravitational potentials of the SMBH and the stellar disk of the axisymmetric model. A triaxial bulge is the primary source of azimuthal angular momentum  within the Galaxy \citep{Gardner:2020jsf}. Our adopted density profile of the triaxial bulge is \citep{1997MNRAS.288..365B, Blg:McGaugh}:
\begin{equation}
    \rho _{\rm Bulge} (x,y,z) = \frac{M_0}{\eta \zeta b_{\rm m} ^3} \ \frac{e^{- ( b/b_{\rm m} )^2}}{\left( 1 + \frac{b}{b_0} \right)^{1.8}} \label{eq:bulge_tri}\, ,
\end{equation}
where
\begin{equation}
    b = \sqrt{x^2 + \frac{y^2}{\eta ^2} + \frac{z^2}{\zeta ^2}}\, ,
\end{equation}
$\eta = 0.5$, $\zeta = 0.6$, $b_{\rm m} = 1.9$ kpc, and $b_0 = 0.1$ kpc \citep{1997MNRAS.288..365B}. The parameters are determined from the observed luminosity distribution \citep{1997MNRAS.288..365B} and indicate that the largest axis of  the bulge, the $x$ axis, is in the plane of the disk, and the second largest axis, the $z$ axis, is in the vertical direction. It follows that two principal axes of the bulge are in the plane of the disk and, therefore, the bulge is not tilted with respect to the plane of the disk. We chose the value of $M_0$ so that the Newtonian accelerations due to the triaxial and the spherical bulges are comparable at $R \gtrsim 0.5$~kpc in the plane of the disk; in other words, the axisymmetric and non-axisymmetric models yield comparable rotation curves beyond $\sim 0.5$~kpc and the dynamical differences between the two models are limited to the central region. We set $M_0 = 5.7 \times 10^{11} M_\sun$, which yields a total mass of the bulge of $2.1 \times 10^{10} M_\sun$. Figure~\ref{fig:bulgedens} compares the density profiles of the spherical and triaxial bulges in our two models. 

We derived the Newtonian gravitational potential of the triaxial bulge, $\Phi_{\rm Bulge}$, by solving the standard Poisson's equation in three dimensions within a box of volume (16 kpc)$^3$ about the center. Figure~\ref{fig:bulgeacc} shows that the magnitudes of the Newtonian radial acceleration for the triaxial bulge are comparable to the Newtonian radial acceleration generated by the spherical bulge of Eq.~(\ref{eq:pot_bulge}) at $R \gtrsim 0.5$~kpc,  as we require with our choice of $M_0$. Within a distance of 0.1 kpc from the center, the spherical bulge generates a larger acceleration due to its steeper density profile shown in Fig.~\ref{fig:bulgedens}. In the intermediate regions between 0.1 kpc and $\sim 0.5$ kpc, the density profile of the spherical bulge is smaller than the density profile of the triaxial bulge; however, the radial acceleration of the spherical bulge is still larger than the acceleration of the triaxial case in any direction because, due to the steeper central density, the spherical bulge encloses a larger mass within $\sim 0.5$ kpc. At large  distances, $r \gtrsim 5$ kpc, the triaxiality is not effective and the mass distribution within the bulge can be treated as spherically symmetric.  

\cite{Blg:McGaugh} investigates the mass model of the MW in the context of MOND by matching the predicted rotation curve against observations. The scale length of the stellar disk and the mass of the bulge determine the relative contributions of these two components of the Galaxy to the total rotation curve. Our model is roughly comparable to the model of \cite{Blg:McGaugh} that has the stellar disk with scale length $R_{\rm d} =$ 4~kpc and bulge mass within 1\% of the bulge mass in our model. Indeed, the circular velocity $v_{\rm circ}$ in the McGaugh model is within 5\% of our $v_{\rm circ}$ within $r=3$ kpc, and $\sim 10\%$ smaller at $r>3$ kpc. For different values of $R_{\rm d}$, the agreement slightly worsens. For example, in McGaugh's model with $R_{\rm d} = 2.3$ kpc,  $v_{\rm circ}$  is  $\sim 30 \%$ smaller than our $v_{\rm circ}$ within $r=3$ kpc, but remains within $\sim 5\%$ at $r>3$ kpc.  

\begin{figure}[!ht]
\centering
\includegraphics[width=3.75cm]{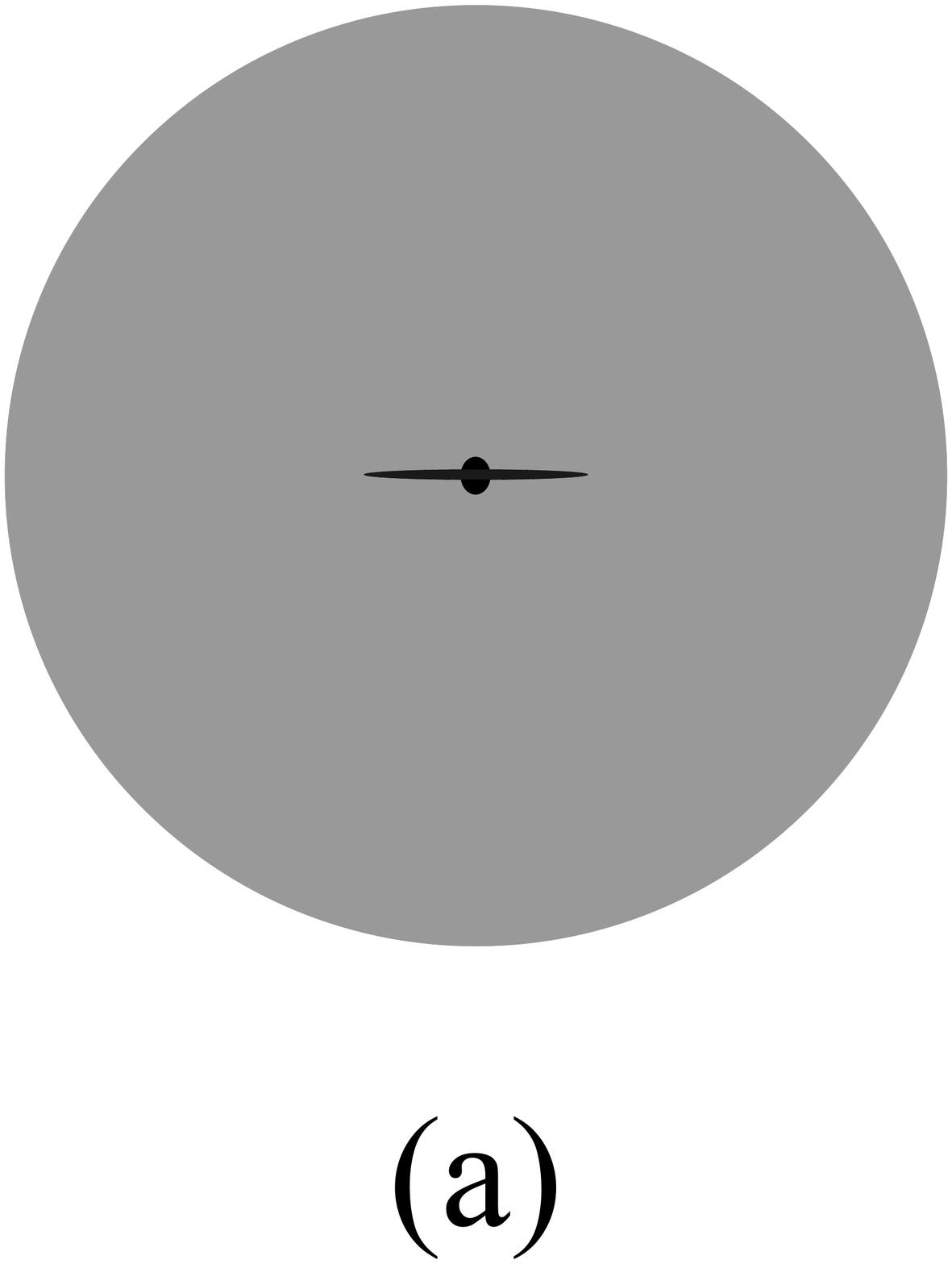}
\includegraphics[width=3.75cm]{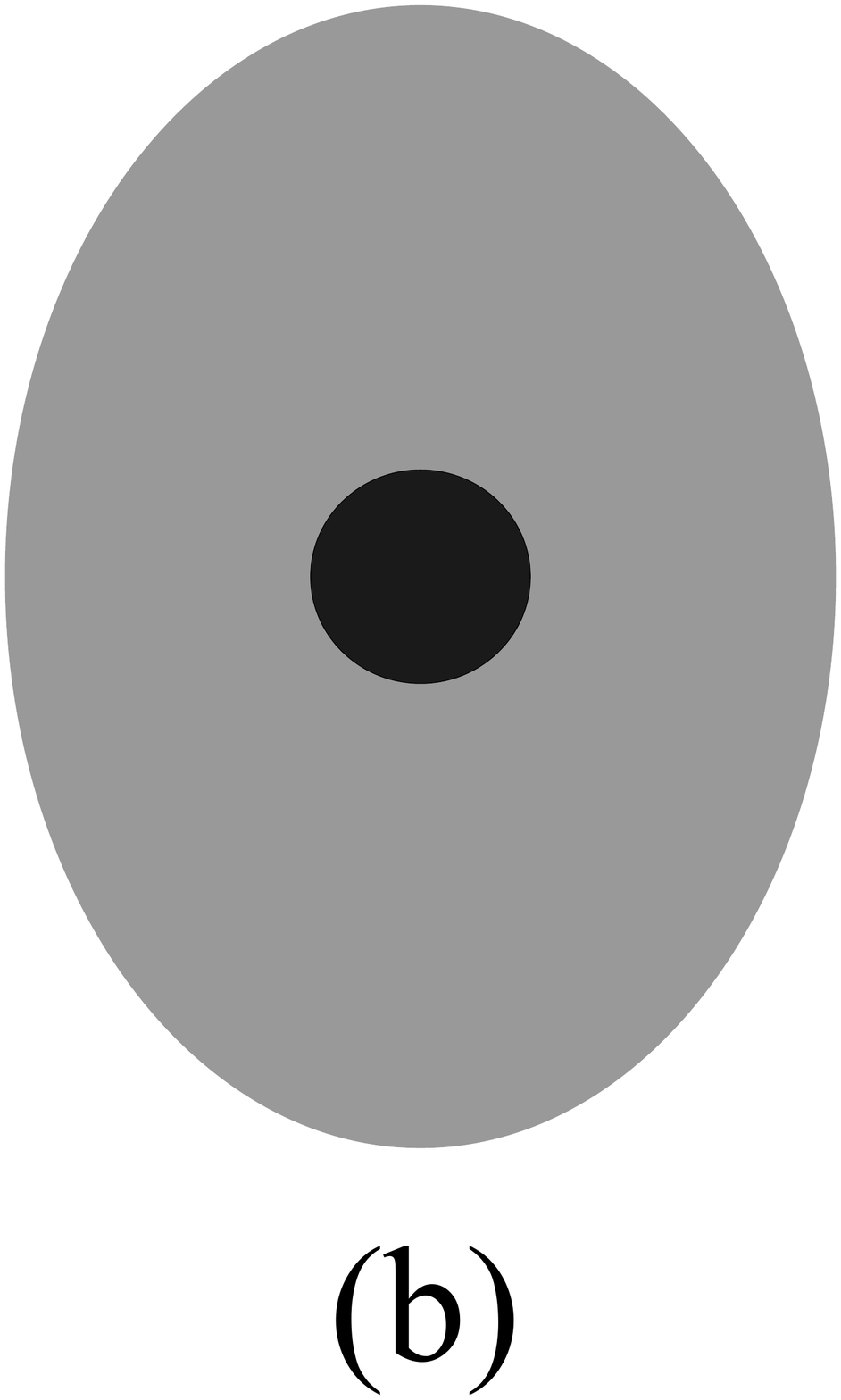}
\includegraphics[width=3.75cm]{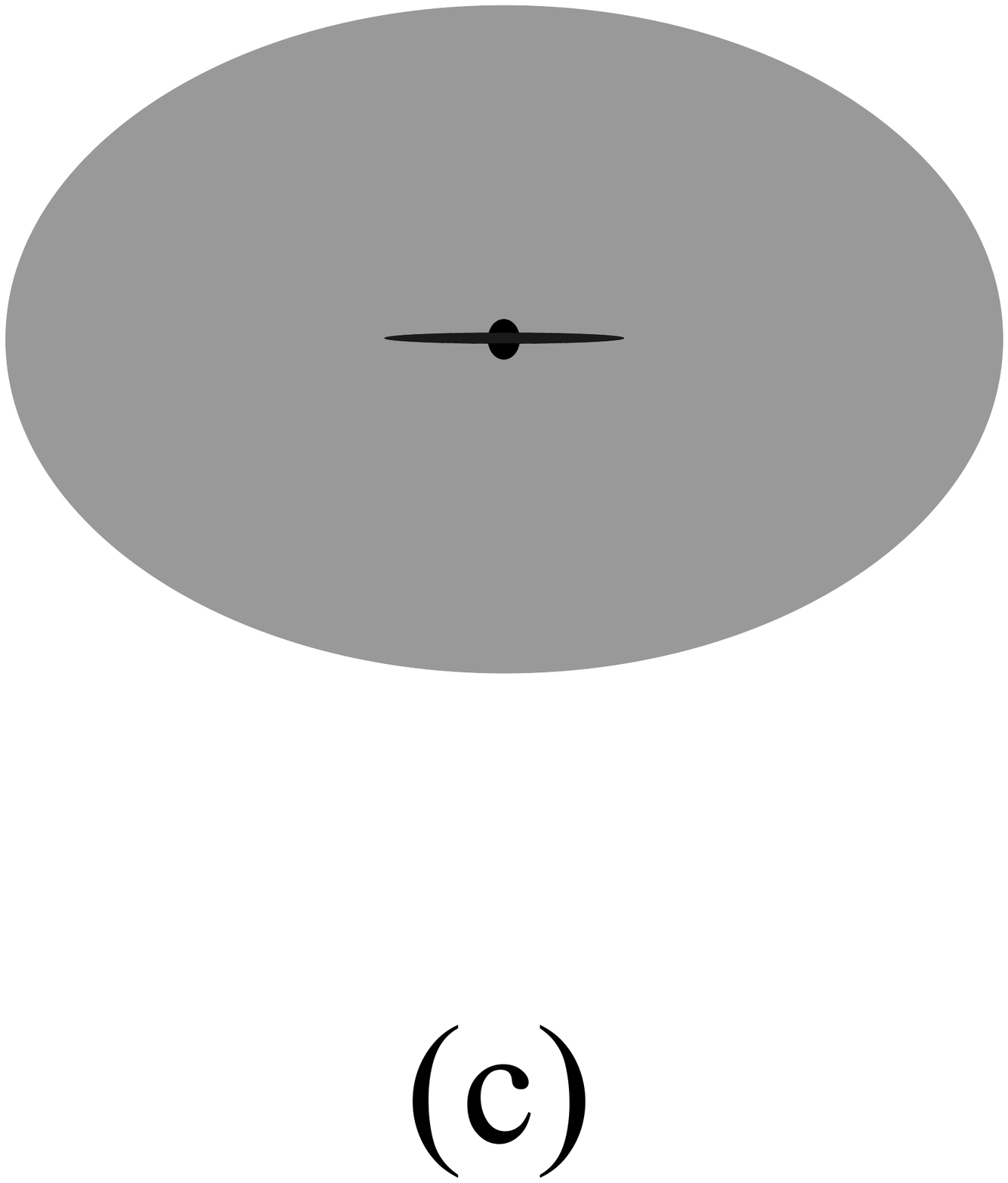}
\includegraphics[width=3.75cm]{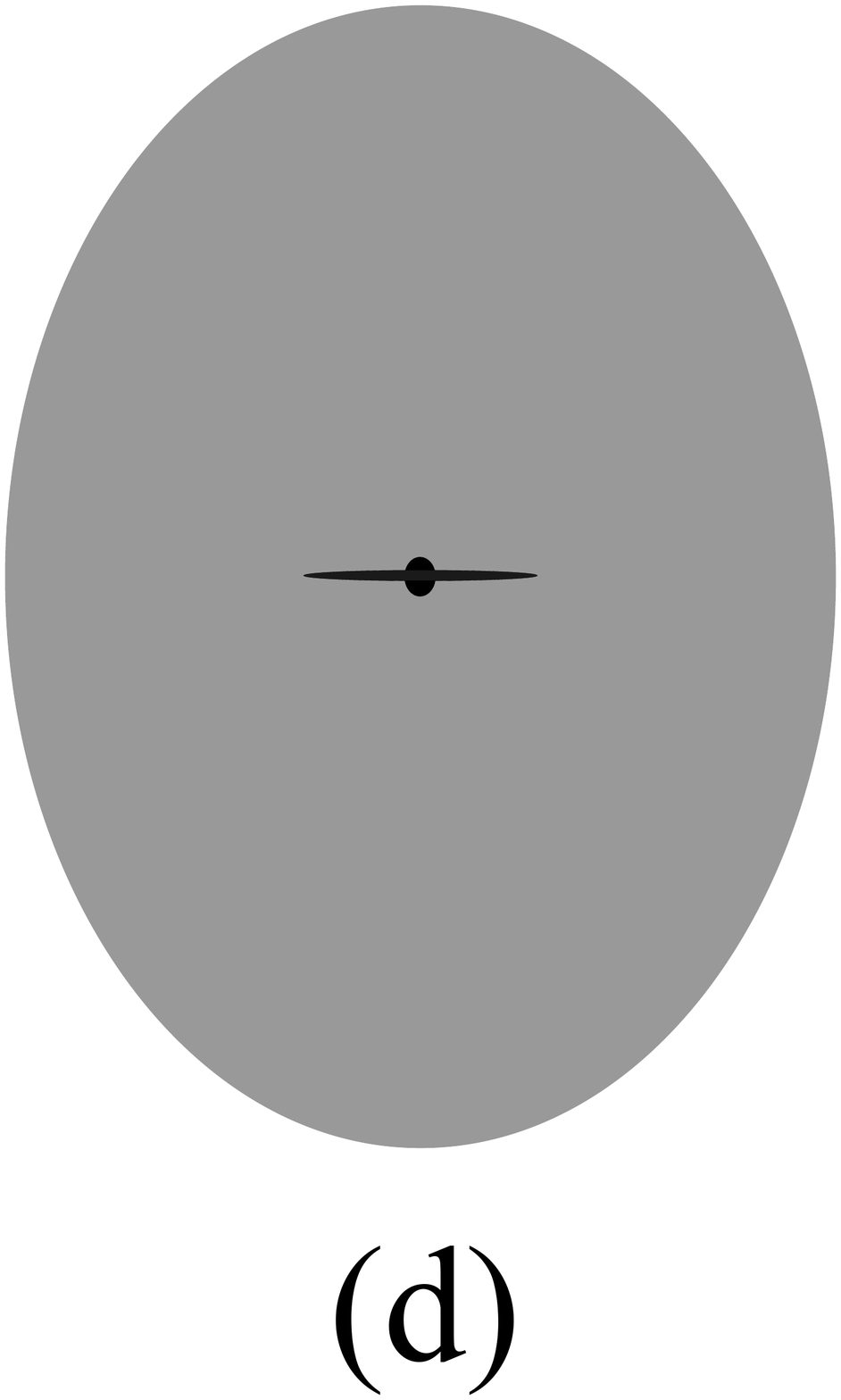}
\caption{\label{fig:MW} Schematic diagrams of our models of the Galaxy in Newtonian gravity. The baryonic components, the bulge and the disk, are shown in black and the dark matter halo in gray. Different shapes of the dark matter halo are shown: (a) spherical (edge-on view of the disk), (b) prolate with the major axis on the plane of disk (face-on view of the disk), (c) oblate with the minor axis perpendicular to the disk (edge-on view of the disk), and (d) prolate with the major axis perpendicular to the disk (edge-on view of the disk). Only in case (b) are the axes of symmetry of the disk and the halo not aligned.}
\end{figure}

\section{The dark matter halo in Newtonian gravity}
\label{sec:dark}

\begin{figure}
\centering
\includegraphics[width=6cm]{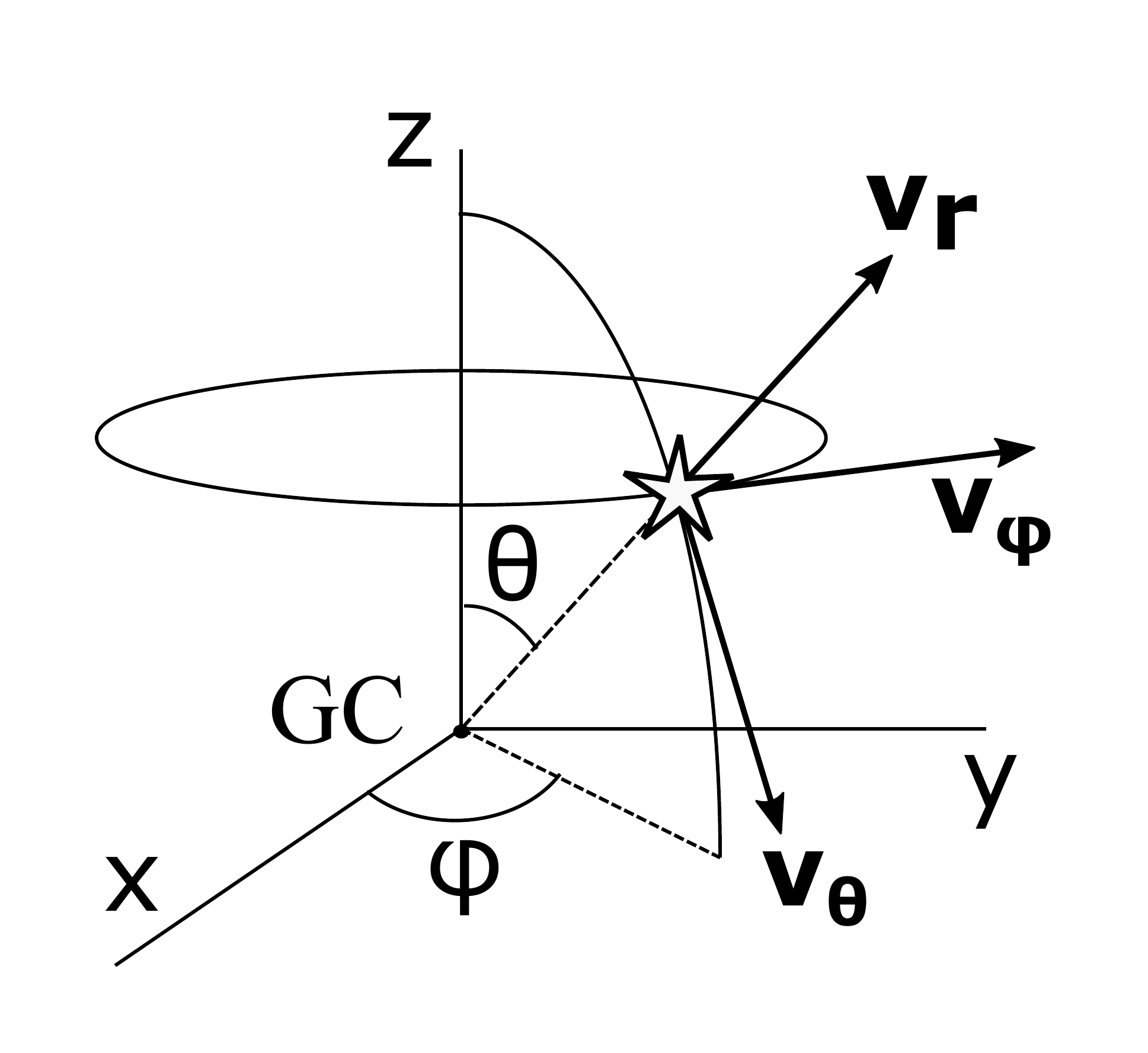}
\caption{\label{fig:vcomp} Components of the velocity of a star in the spherical polar coordinates with origin at the Galactic center (GC): $\vec v = \vec v_r + \vec v _\theta + \vec v _\phi$.  }
\end{figure}

We needed to compare the dynamics of HVSs in QUMOND with the corresponding expectations in Newtonian gravity. We thus also considered a Newtonian model of the MW with the same axisymmetric baryonic components described in the previous section, but with an additional surrounding dark matter halo.

For the dark matter halo, we adopted the triaxial generalization suggested by \citet{Halo:Vogelsberger} of the spherical Navarro-Frenk-White (NFW) gravitational potential \citep{Halo:NFW}: 
\begin{equation}
\Phi_{\rm Halo} = - \frac{G M_{\rm 200}}{f(C_{\rm 200})} \ \frac{1}{\tilde{r}} \ \ln \left( 1 + \frac{\tilde{r}}{r_{\rm s}} \right) \ , \label{eq:pot_halo}
\end{equation}
where $f(u) = \ln (1 + u) - u/(1+u)$. $M_{\rm 200} = 8.35 \times 10^{11} M_\sun$ is the mass within $r_{\rm 200}$,\footnote{$r_{\rm 200}$ is the radius of a spherical volume within which the mean mass density is 200 times the critical density of the Universe.} $C_{\rm 200} = r_{\rm 200}/r_{\rm s} = 10.82$ is the concentration parameter, and $r_{\rm s} = 18$ kpc is the scale radius. Our adopted values of the parameters are those used in \cite{Halo:Hesp}. They are consistent with the estimates from the kinematics of halo stars \citep{2008ApJ...684.1143X, 2012MNRAS.425.2840D}. The generalized radius is
\begin{equation}
\tilde{r} = \frac{r_E (r + r_{\rm a})}{r_E + r_{\rm a}}\, ,
\label{eq:rtilde}
\end{equation}
where the ellipsoidal radius $r_E$ is 
\begin{equation}
r_E = \sqrt{\left( \frac{x}{a} \right)^2 + \left( \frac{y}{b} \right)^2 + \left( \frac{z}{c} \right)^2}\, 
\end{equation}
and $r_{\rm a} = 1.2 r_{\rm s}$ \citep{Halo:Hesp} is the length scale that determines the transition from the triaxial to the spherical shape: In the inner region ($r \ll r_{\rm a}$) the halo is triaxial ($\tilde{r} \approx r_E$), whereas in the outer region ($r \gg r_{\rm a}$) the halo is almost spherical ($\tilde{r} \approx r$). This transition in the shape of the halo is a generic prediction of $\Lambda$CDM simulations (see, e.g., \citealt{2007MNRAS.377...50H}). The parameters $a$, $b$ and $c$ satisfy the relation $a^2 + b^2 + c^2 = 3$. 

We defined the two triaxiality parameters $q_y = b/a$ and $q_z = c/a$. Once $q_y$ and $q_z$ are specified, $a$ is given by 
\begin{equation}
    a = \sqrt{\frac{3}{1+ q_y ^2 + q_z ^2}} \ \ . \label{eq:a}
\end{equation}
Since the shape of the halo is currently poorly constrained  \citep{MW:review}, we varied either $q_y$ or $q_z$ within 40\% from  unity. 

In this paper we did not use a triaxial dark matter halo. We only explored spheroidal halos with different axes of symmetry (see Fig.~\ref{fig:MW}). To study the effects of the dark matter halo shape on the azimuthal component of the HVS velocity, $v_\phi$ (see Fig.~\ref{fig:vcomp}), we chose a spheroidal halo that is axisymmetric about the $y$ axis lying in the plane of the disk (case (b) of Fig.~\ref{fig:MW}) whereas the baryonic components are axisymmetric about the vertical $z$ axis. In this model, we set $q_y > 1$ and $q_z = 1$. To study the effects of the dark matter halo shape on the latitudinal component $v_\theta$ of the HVS velocities, we considered a spheroidal halo that is axisymmetric about the $z$ axis (i.e., $q_y = 1$ and $q_z \neq 1$). The halo is oblate if $q_z < 1$ and prolate if $q_z > 1$ (cases (c) and (d) of Fig.~\ref{fig:MW}). Indeed, the kinematics of halo stars from the Sloan Digital Sky Survey suggests that the dark matter halo might be oblate within $\sim 20$~kpc, with $q_z=0.7\pm 0.1$, based on the estimate of constant potential surfaces \citep{Loebman_2014}.

At the solar neighborhood ($R = 8$ kpc), our models yield a circular velocity of 235 km~s$^{-1}$ in Newtonian gravity and 240 (210) km~s$^{-1}$ in QUMOND with $\gamma = 1$ ($\gamma = 2$). In addition, in our Newtonian model, the total mass enclosed within 120 pc is in perfect agreement with the observed value reported in Table 3 of \cite{kenyon2008}, derived from the estimate of the  mass of the central region of the stellar disk of \cite{2002A&A...384..112L}. The MW central region is where the HVSs experience the largest deceleration. We verified that varying the radial acceleration profile of the MW central region within the observational uncertainties does not affect our conclusions on the tangential  velocities of the HVSs.

\section{Simulations of the kinematics of the HVSs}
\label{sec:HVS}

In this section we describe our simulations and our synthetic samples of HVSs. Section \ref{sec:simulation} illustrates the initial conditions of the equations of motion of the HVSs that we adopted. In Sect.~\ref{sec:sample} we show that only stars beyond a minimum galactocentric distance, corresponding to a minimum ejection velocity, are relevant to discriminate between QUMOND and Newtonian gravity. This threshold depends on the star mass: for $4~M_\sun$ stars, the minimum galactocentric distance is 15 kpc, which corresponds to the minimum ejection velocity $\sim 750$ km~s$^{-1}$ for models in Newtonian gravity and QUMOND with $\gamma = 1$. For QUMOND with $\gamma = 2$, the ejection velocity threshold is $\sim 710$ km~s$^{-1}$, which also corresponds to a minimum distance of 15 kpc.

\subsection{Simulation setup}
\label{sec:simulation}

We assumed that the HVSs are generated through Hills' mechanism \citep{Hills88}: In a three-body interaction between the SMBH associated with SgrA$^{\star}$ and a binary star, one star of the binary is ejected and the other is captured by the black hole. We emphasize that the assumption of Hills' mechanism for the ejection of the HVSs does not affect our conclusions, as detailed at the end of this section.

In Hills' ejection scenario, we simulated the velocity distribution of the HVSs ejected from the Galactic center by means of a three-body numerical code that reproduces the encounter of a set of equal-mass binary stars with the SMBH. The binaries' orbital parameters and minimum approach distance to the SMBH are drawn from appropriate distributions \citep[][and references therein]{bromley2006}. More details on the numerical code and on the ejection velocity distribution will be provided in \citet{Gallo} and in an additional, separate paper. Here, we only report the information that is instrumental to the present analysis. 

We adopted a mass $M_{\rm BH} = 4 \times 10^6 M_\sun$ for the black hole (as in our model in Sect. \ref{sec:axisymm}) and a mass of 4~$M_\sun$ for each star in the binary. 
The choice of mass for the binary stars is consistent with the fact that most of the observed unbound HVSs have masses between 2.5 and 4 $M_\sun$ \citep{HVS:review}. In addition, choosing the upper limit of this mass range is a conservative choice that is appropriate for our investigation, as we clarify in Sect.~\ref{sec:angvel}. We obtained a distribution of ejection speeds that displays a prominent peak at $v_{\rm ej} \sim 510$~km~s$^{-1}$, extends to velocities of $\sim 4000$~km~s$^{-1}$, and has a positive skewness; 16\% (12\%) of the ejected stars have speed $v_{\rm ej}\gtrsim 710$ (750) km~s$^{-1}$, a value that will be relevant for the present analysis (see Sect. \ref{sec:sample}). Our results on the ejection velocities are comparable to the results obtained from the analytical prescriptions provided by \citet{bromley2006}.

The distribution of ejection speeds formally corresponds to the distribution of velocities that the ejected stars would have at infinite distance from the SMBH in absence of other sources of gravitational potential. However, in the context of the Galactic center, this distribution can be taken as the velocity distribution of the ejected stars at the starting point of their trajectories across the MW, which we set as the radius of the sphere of influence of the SMBH \citep[i.e.,\ $r = 3$ pc;][]{genzel2010}. Although the distribution of ejection velocities weakly depends on the binary mass \citep{bromley2006}, our results regarding the kinematics of the HVSs are independent of their mass, because the gravitational acceleration acting on the HVSs is mass-independent.

\begin{figure*}[!ht]
\centering 
\includegraphics[width=5.8cm]{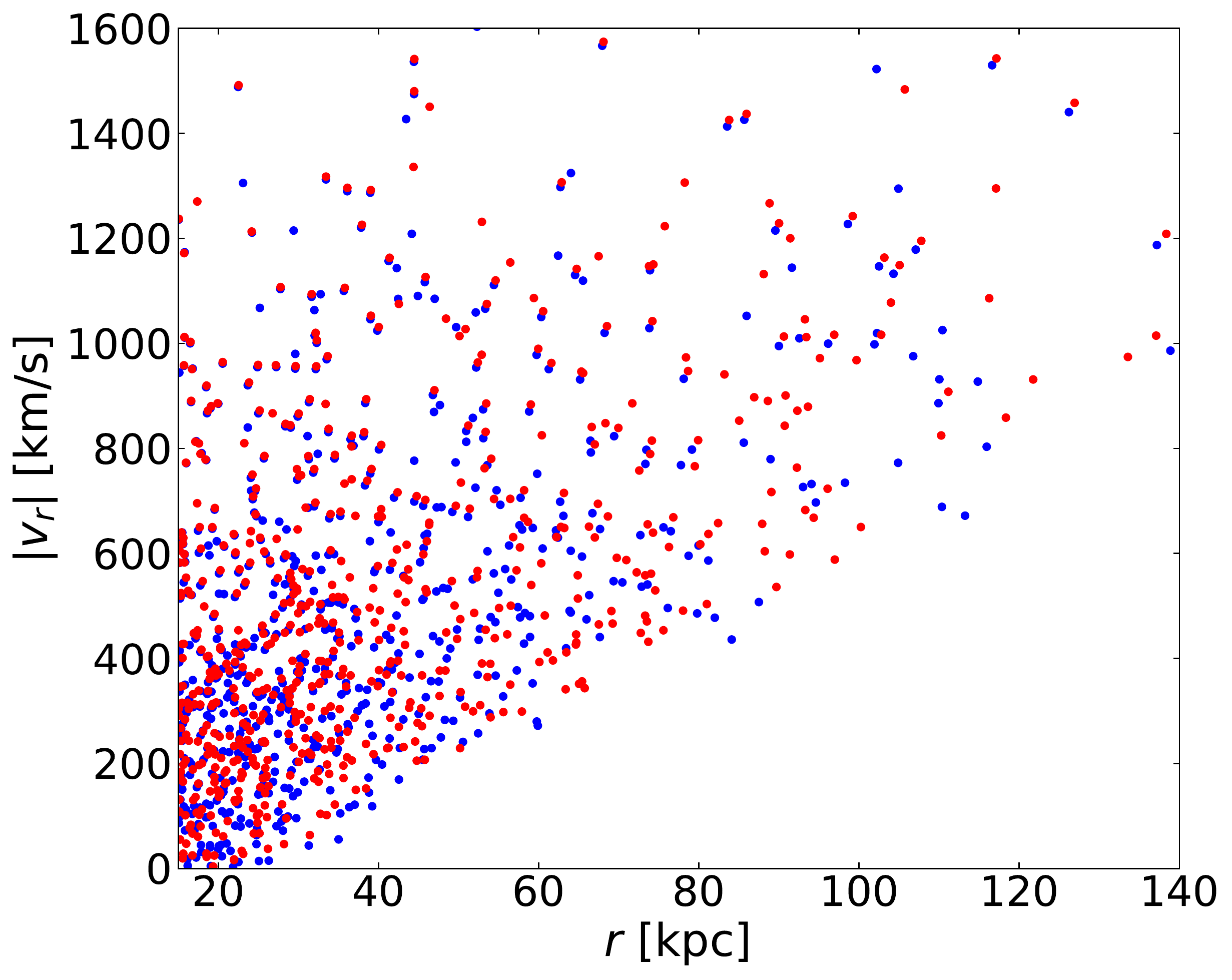}
\includegraphics[width=5.8cm]{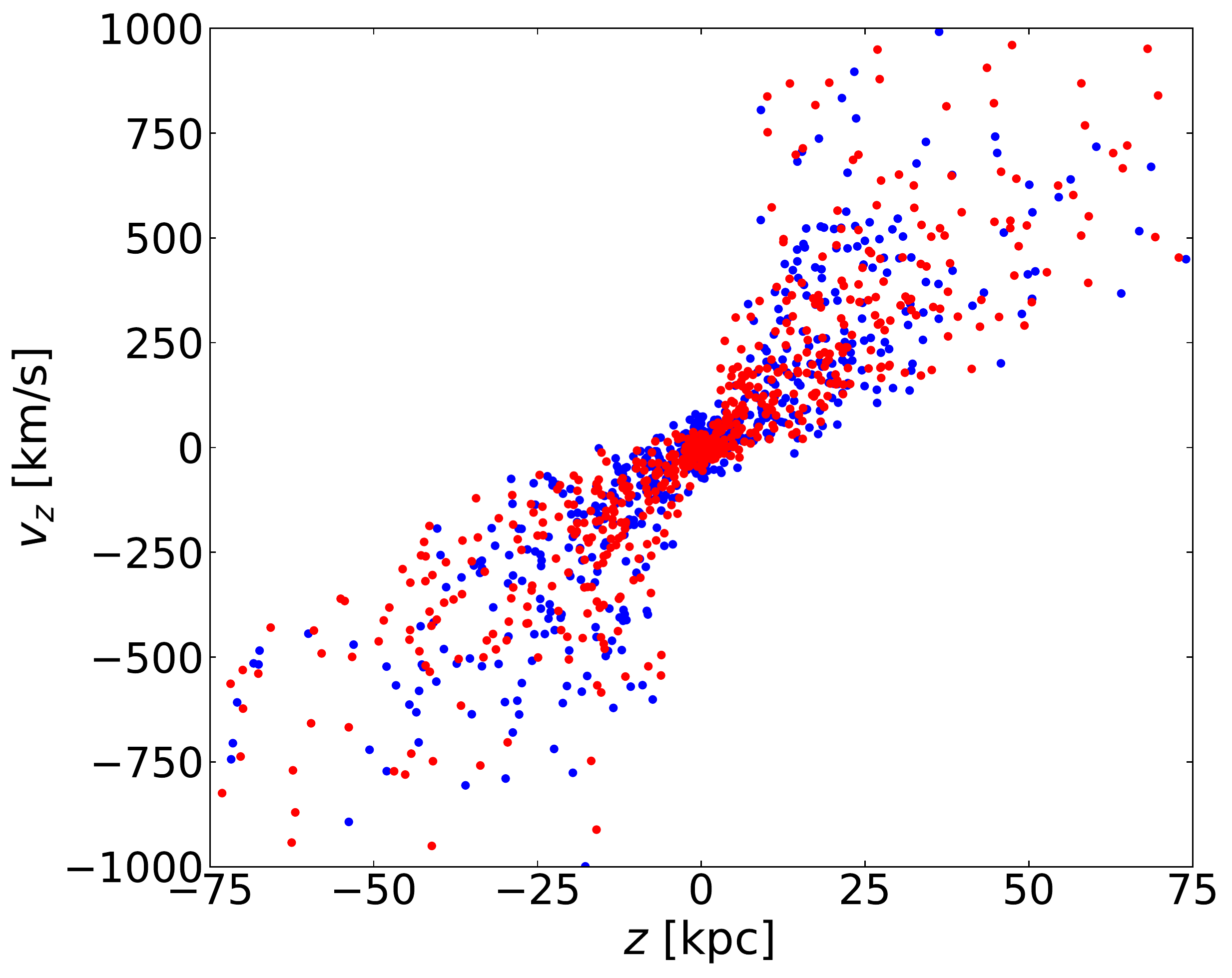}
\includegraphics[width=5.8cm]{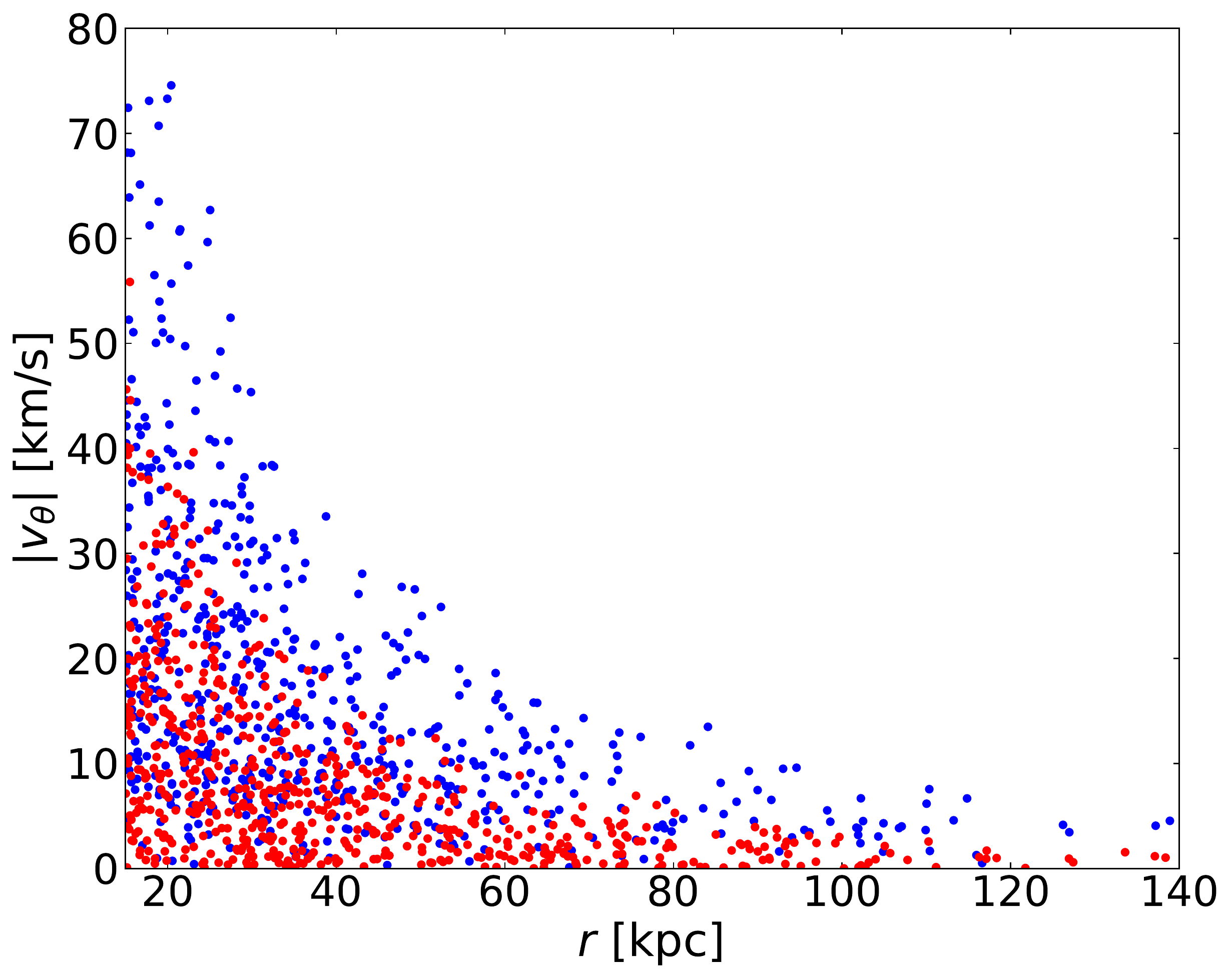}
\caption{\label{fig:compare} Distributions of HVSs in three sections of phase space in QUMOND with $\gamma = 1.0$ (blue dots) and Newtonian gravity with a prolate ($q_y = 1, q_z = 1.1$) dark matter halo (red dots). The left, middle, and right panels show the $r$-$v_r$, $z$-$v_z$, and $r$-$v_\theta$ sections, respectively. Here, $r$ is the galactocentric distance, $z$ is the vertical coordinate, and $v_r$, $v_z$, $v_\theta$ are the radial, vertical, and latitudinal components of the velocity, respectively. }
\end{figure*}

The direction $\hat{n} (\theta, \phi)$ of the ejection velocity is assigned randomly to each star, because Hills' mechanism yields isotropic ejections. We used
the fourth order Runge-Kutta method with adaptive step-size to integrate the equation of motion of each star. We started with a predefined time-step of 5 kyr and the time-step was adjusted so that, at each step, the position and velocity of the star were determined with an accuracy of 1 pc and 0.01 km~s$^{-1}$, respectively. For stars with ejection velocities higher than 600 km~s$^{-1}$, the conservation of total energy holds with a relative accuracy of $10^{-12}$ for each time-step as well as over the entire simulation. Stars with  ejection velocities lower than 600~km~s$^{-1}$ are not relevant for our study.

The typical lifetime of an isolated star of 4~$M_\sun$ on the main sequence is $\tau_{\rm L} \simeq 160$ Myr, for Solar metallicities \citep{schaller1992,brown2006}. We assumed this lifetime as the total lifetime of our simulated stars. At the time of ejection, each star was assigned a random age of $\tau_{\rm ej} = \epsilon \tau_{\rm L}$, where $\epsilon$ is a random number drawn from a uniform density distribution between 0 and 1. We took the average ejection rate to be $10^{-4} \ {\rm yr} ^{-1}$, which is consistent with the estimates by \cite{YuTremaine} and \cite{2013ApJ...768..153Z} \citep[see also][]{Hills88}. Therefore, in the simulation, we ejected stars in intervals of $\Delta t = 0.01$ Myr. For our chosen lifetime and ejection rate, the distribution of ejected stars in the MW reaches a steady state after 160 Myr. We started the simulation at $t=0$ and the $i$-th star was ejected at $t = (i-1) \Delta t$ with its age $\tau^i _{\rm ej}$ at the time of ejection. We chose the time of observation, which is the total run-time of the simulation, to be $t_{\rm obs}=400$ Myr when the distribution of the HVSs is in the stationary regime. For each star, we 
tested the condition of its survival: A star survives long enough to be observed if $[t_{\rm obs} - (i-1) \Delta t] < (\tau_{\rm L} - \tau^i _{\rm ej}) $. If a star satisfies this condition, its travel time is determined by $[t_{\rm obs} - (i-1) \Delta t]$.

We simulated the kinematics of the HVSs through the Galaxy (i) for QUMOND, with the baryonic components only, and (ii) for Newtonian gravity, with the baryonic components and the dark matter halo. For each model of the Galactic potential, we determined positions and velocities of the HVSs at the time of observation, $t = t_{\rm obs}$. Tables ~\ref{table:QUMOND} and \ref{table:Newtonian} summarize the simulations that we performed in QUMOND and in Newtonian gravity. The tables list both the simulations of our simpler model for the baryonic Galactic components considered in Eq.~(\ref{eq:gN}) and described in Sect.~\ref{sec:pot} and  the  simulations that include the additional nonspherical HG halo surrounding the MW (see Sect.~\ref{sec:HG}).

\begin{table}
\caption{List of the simulations for QUMOND.}
\label{table:QUMOND} 
\centering  
\begin{tabular}{c c c c} 
\hline\hline
$\gamma$ & Bulge & Hot gaseous halo & Number of stars \\ 
\hline          
    1 & spherical & no & 40000 \\  
    1 & triaxial & no & individual stars\\ 
    2 & spherical & no & 40000 \\ 
    2 & triaxial & no & individual stars \\ 
    1 & spherical & yes & 40000\\ 
    2 & spherical & yes & 40000 \\ 
\hline
\end{tabular}
\tablefoot{To find the upper limit of $\vert v_\phi \vert$ for QUMOND with a triaxial bulge, we performed many simulations of individual stars. In this case, we used the gravitational field of the triaxial bulge within 5 kpc of the Galactic center and that of the spherical bulge at larger radii.}
\end{table}

\begin{table}
\caption{List of the simulations for Newtonian gravity.} 
\label{table:Newtonian} 
\centering  
\begin{tabular}{c c c c} 
\hline\hline
$q_y$ & $q_z$ & Panel of Fig.~\ref{fig:MW} & Hot gaseous halo \\ 
\hline          
    1 & 0.9 & c & no \\  
    1 & 1 & a & no \\ 
    1 & 1.1 & d & no \\ 
    1.1 & 1 & b & no \\ 
    1.2 & 1 & b & no \\ 
    1.2 & 1 & b & yes \\
    1.4 & 1 & b & yes \\
\hline
\end{tabular}
\tablefoot{Simulations in Newtonian gravity were performed with a spherical bulge. Each simulation contains 40000 stars.}
\end{table}

We compared the final phase-space distributions of the HVSs obtained in QUMOND  with the distributions obtained in Newtonian gravity. Due to the large radial velocities of the HVSs, their phase-space distributions in the $(r, v_r)$ or  $(z, v_z)$ space in QUMOND  are virtually indistinguishable from those in Newtonian gravity (left and middle panels of Fig.~\ref{fig:compare}). However, the distributions of the galactocentric tangential velocity components, the latitudinal component $v_\theta = r ({\mathrm{d}}\theta/{\mathrm{d}}t)$, and the azimuthal component $v_\phi = r \sin \theta ({\mathrm{d}}\phi/{\mathrm{d}}t)$, are  distinctive and can be used, in principle, to discriminate between QUMOND and Newtonian gravity (right panel of Fig.~\ref{fig:compare}). Our result is independent of the mechanism responsible for the ejection of the HVSs from the Galactic center: any mechanism that can expel stars from the Galactic center with radial velocities $v_{\rm ej}\gtrsim 710$ km~s$^{-1}$, as we discuss below, and null tangential velocities would be suitable to perform the analysis presented in this work, leading to the same conclusions.

\subsection{Evolution of the tangential velocity components}
\label{sec:sample}

The tangential velocity components of the HVSs are excellent probes of the nonspherical components of the Galactic gravitational potential. However, this distinctive ability only holds for HVSs with ejection speeds higher than a threshold: for stars with lower ejection speed, the tangential velocity components may indeed be disproportionately high.

To find the threshold ejection velocity, we simulated the dynamics of 200 HVS with mass 4~$M_\sun$, ejection velocities $v_{\mathrm {ej}}$ between 650 and 850 km~s$^{-1}$, and random initial directions. For each star, the simulation time was taken to be its lifetime, namely 160 Myr for 4~$M_\sun$ stars. 
This time is the maximum possible travel time. For each star, we determined the maximum values of the magnitudes of the tangential velocity components, $|v_\theta|$ and $|v_\phi|$.

We considered both Newtonian gravity with different shapes of the dark matter halo and  QUMOND with $\gamma = 1$ and $\gamma=2$. For both the Newtonian and the QUMOND scenarios, the baryonic components (Eqs.~\ref{eq:pot_disk} - \ref{eq:pot_BH}) were taken to be axisymmetric about the $z$ axis. Due to the pull of the disk, the stars obtain nonzero $v_\theta$ values for all the models. However, in QUMOND with a spherical bulge, the stars do not obtain any $|v_\phi|$ because the baryonic matter, the only component present, is axisymmetric. In our models of Newtonian gravity, the stars acquire nonzero values of $v_\phi$ only when the two axes of symmetry of the baryonic and dark matter components are misaligned: for example, for $(q_y, q_z) = (1.1, 1)$ (panel (b) in Fig.~\ref{fig:MW}), the dark halo is symmetric about the $y$ axis whereas the baryonic matter is symmetric about the $z$ axis. When no such misalignment is present, as in the cases sketched in panels (a), (c), and (d) of Fig. \ref{fig:MW}, $v_\phi$  vanishes, and the only non-null component of the tangential velocity is the latitudinal velocity, $v_\theta$.
 
\begin{figure*}[!h]
\centering
\includegraphics[width=8.5cm]{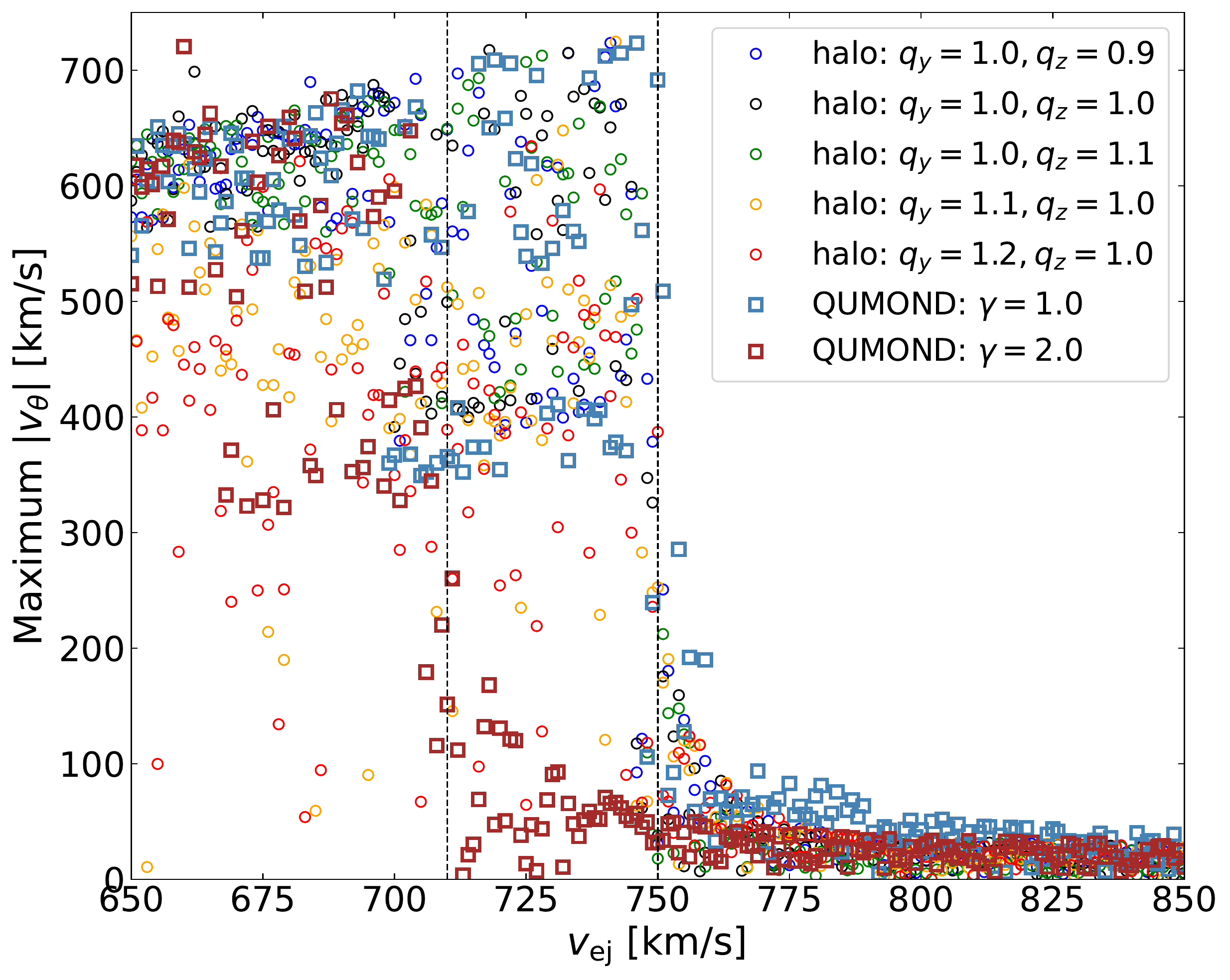}
\includegraphics[width=8.5cm]{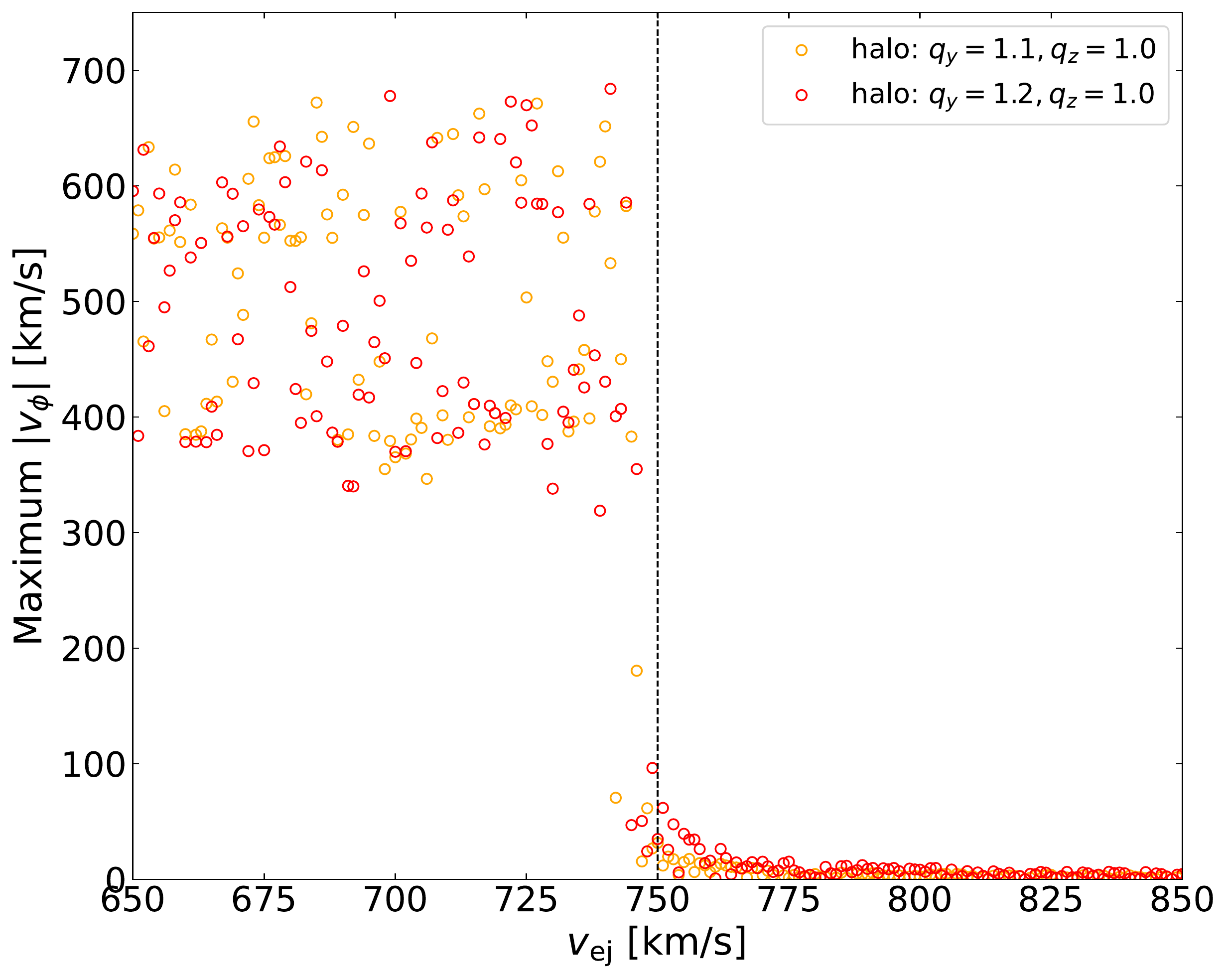}
\caption{ {\it Left panel:} Maximum values of the latitudinal velocity components, $|v_\theta|$, of 4~$M_{\sun}$ HVSs as a function of their ejection velocities, $v_{\rm ej}$,  in Newtonian gravity with different shapes of the dark matter halo and in QUMOND with $\gamma = 1$ and $\gamma=2$; symbols and models are detailed in the inset. The vertical scatter in the plot originates from the random directions of the star ejection. The vertical dotted lines indicate $v_{\rm ej}= 710$ and $v_{\rm ej}=750$ km s$^{-1}$.  In all the models except QUMOND with $\gamma = 2$, the maximum $|v_\theta|$ substantially drops for stars with $v_{\rm ej}\gtrsim 750$ km s$^{-1}$. For QUMOND with $\gamma = 2$, the maximum $|v_\theta|$ drops for stars with  $v_{\rm ej}\gtrsim  710$ km s$^{-1}$. {\it Right panel:} Maximum values of the azimuthal velocity components, $|v_\phi|$, of 4~$M_{\sun}$ HVSs as a function of their ejection velocities, $v_{\rm ej}$,  in two Newtonian gravity models where the axes of symmetry of the baryonic matter and of the dark matter halo are misaligned.  
The other models considered in the left panel do not appear because they are axisymmetric and have zero $|v_\phi|$ values. The vertical scatter originates from the random directions of the star ejection. The vertical dotted line indicates $v_{\rm ej}=750$ km s$^{-1}$. }
\label{fig:vejvphimax}
\end{figure*}

The left panel of Fig.~\ref{fig:vejvphimax} shows that $|v_\theta|$ can be as large as 600 km~s$^{-1}$ for stars with $v_{\rm ej} < 710$ km~s$^{-1}$ in QUMOND with $\gamma = 2$,  and for stars with $v_{\rm ej} < 750$ km~s$^{-1}$ in all the other models. In contrast, for stars with ejection speed higher than these thresholds, the maximum values of $|v_\theta|$  are consistently lower than $\sim 100$ km~s$^{-1}$. The right panel of Fig.~\ref{fig:vejvphimax} shows qualitatively similar results for $|v_\phi|$ in the Newtonian models with the misaligned axes of symmetry ($q_y \neq 1$): in these models, $|v_\phi|$ is lower than $\sim 20$ km~s$^{-1}$ when $v_{\rm ej} \gtrsim 750$ km~s$^{-1}$. In these Newtonian models with $q_y \neq 1$, the vertical scatter in the maximum values of $|v_\theta|$ (left panel) is large. The stars ejected with smaller angle with respect to the disk (i.e., $(90^\circ - \theta) \lesssim 30^\circ$) undergo larger deflection and attain larger $|v_\theta|$. On the other hand, the stars ejected almost perpendicular to the disk do not bend significantly, hence have smaller $|v_\theta|$. Conversely, in the same Newtonian models, the stars always attain a $|v_\phi| \gtrsim 350$ km~s$^{-1}$ (right panel) irrespective of the direction of the HVS ejection velocity.

\begin{figure}[!ht]
\centering
\includegraphics[width=8cm]{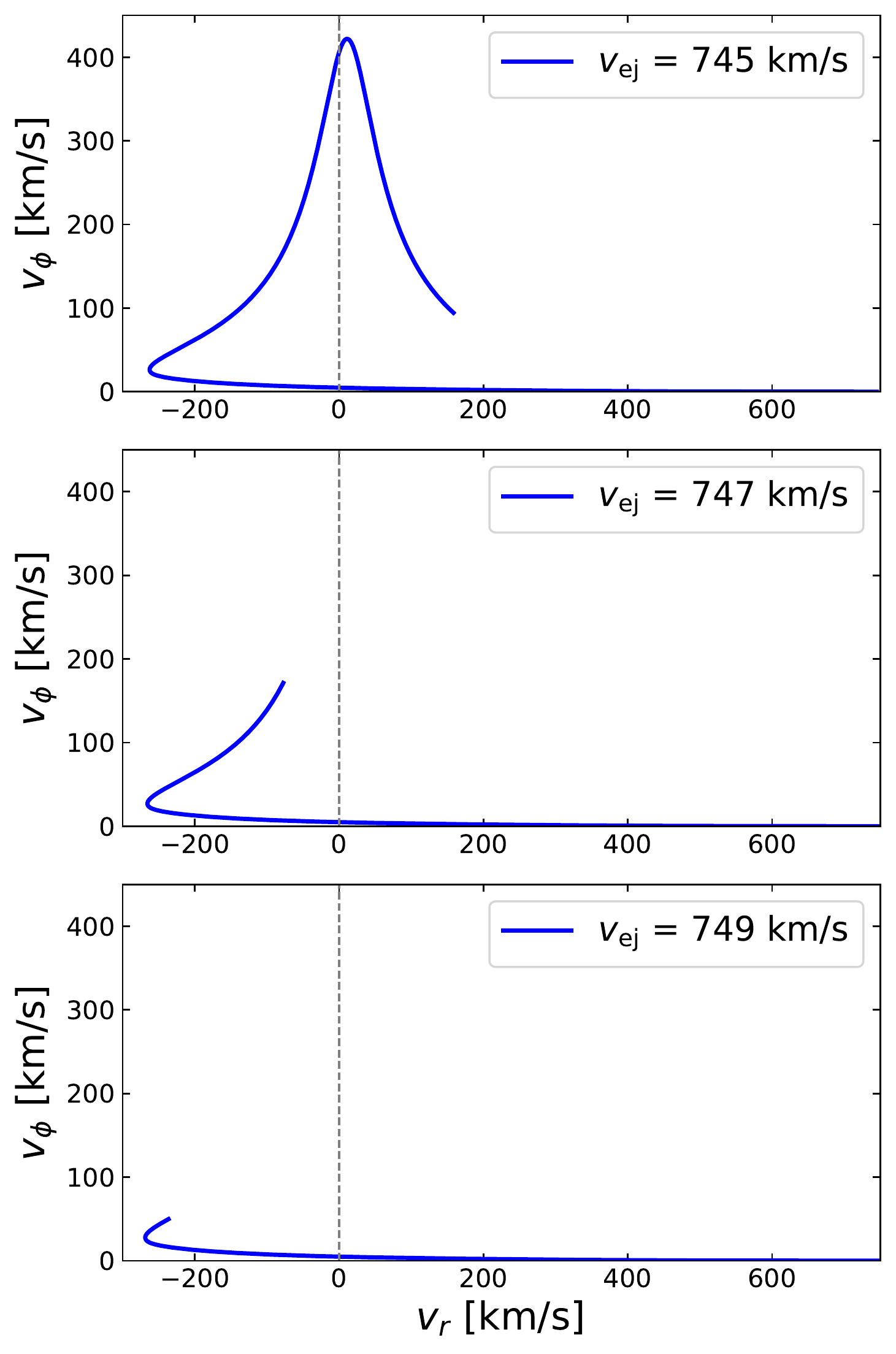}
\caption{\label{fig:oneHVS} Azimuthal velocity component, $v_\phi$, as a function of the radial velocity component, $v_r$, for three HVSs with mass 4~$M_{\sun}$ and ejection velocity $v_{\rm ej} = \{745, 747, 749\}$ km~s$^{-1}$ in Newtonian gravity with $(q_y, q_z) = (1.1, 1)$. In each case, the star starts with $v_r = v_{\rm ej}$ and $v_\phi=0$. The maxima of $v_\phi$ are quite different in the three cases. The stars with $v_{\rm ej} = 747$ and 749 km~s$^{-1}$ do not live long enough to undergo the inner turnaround, that is, they do not encounter $v_r = 0$ for the second time. On the contrary, the inner turnaround occurs for the star with $v_{\rm ej} = 745$ km~s$^{-1}$.}
\end{figure}

High tangential velocities of the stars with lower ejection speed are caused by the exchange of kinetic energy between radial and angular degrees of freedom, especially when the stars undergo an inner turnaround. Figure~\ref{fig:oneHVS} illustrates that the inner turnaround does occur for stars with low ejection velocity. In Fig.~\ref{fig:oneHVS} we show $v_\phi$ as a function of the radial component $v_r$ for three stars with ejection velocity $v_{\rm ej} = \{745, 747, 749\}$ km~s$^{-1}$ in the Newtonian models with the misaligned axes, with $(q_y, q_z) = (1.1, 1)$. In each case, $v_r = v_{\rm ej}$ and $v_\phi = 0$, initially; at later times, $v_\phi$ starts increasing as $v_r$ decreases. When $v_r = 0$ for the first time, as indicated  by the vertical dotted lines, the star undergoes the outer turnaround, namely it reaches its maximum distance from the MW center and starts moving inward. Therefore, $v_r$ becomes negative while $v_\phi$ keeps increasing. When $v_r$ becomes zero for the second time, the star undergoes the inner turnaround, namely the star reaches the closest approach to the center and starts again moving outward. Around this phase of the inner turnaround, $|v_\phi|$ reaches its maximum, as shown for $v_{\rm ej} = 745$ km~s$^{-1}$. Stars with higher ejection speeds take longer time to undergo the  outer turnaround and may not live long enough to experience the inner turnaround, as illustrated by the other two stars shown in Fig.~\ref{fig:oneHVS}. As a result, the maximum of $|v_\phi|$ falls sharply around $v_{\rm ej} \approx 750$ km~s$^{-1}$ (right panel of Fig.~\ref{fig:vejvphimax}).

We conclude that for stars with ejection velocities higher than a threshold velocity, the maximum values of the magnitudes of the latitudinal and azimuthal velocity components, $|v_\theta|$ and $|v_\phi|$, are proportionate to the departure from the spherical symmetry of the potential. For stars with ejection speeds lower than the threshold, this proportionality disappears because these stars may experience the inner turnaround. Our simulations suggest that, for 4 $M_{\sun}$ stars, this threshold ejection velocity is $\sim 710$ km~s$^{-1}$ for QUMOND with $\gamma = 2$ and $\sim 750$ km~s$^{-1}$ for all the other models we investigate. 

\begin{figure}[!ht]
\centering
\includegraphics[width=8.5cm]{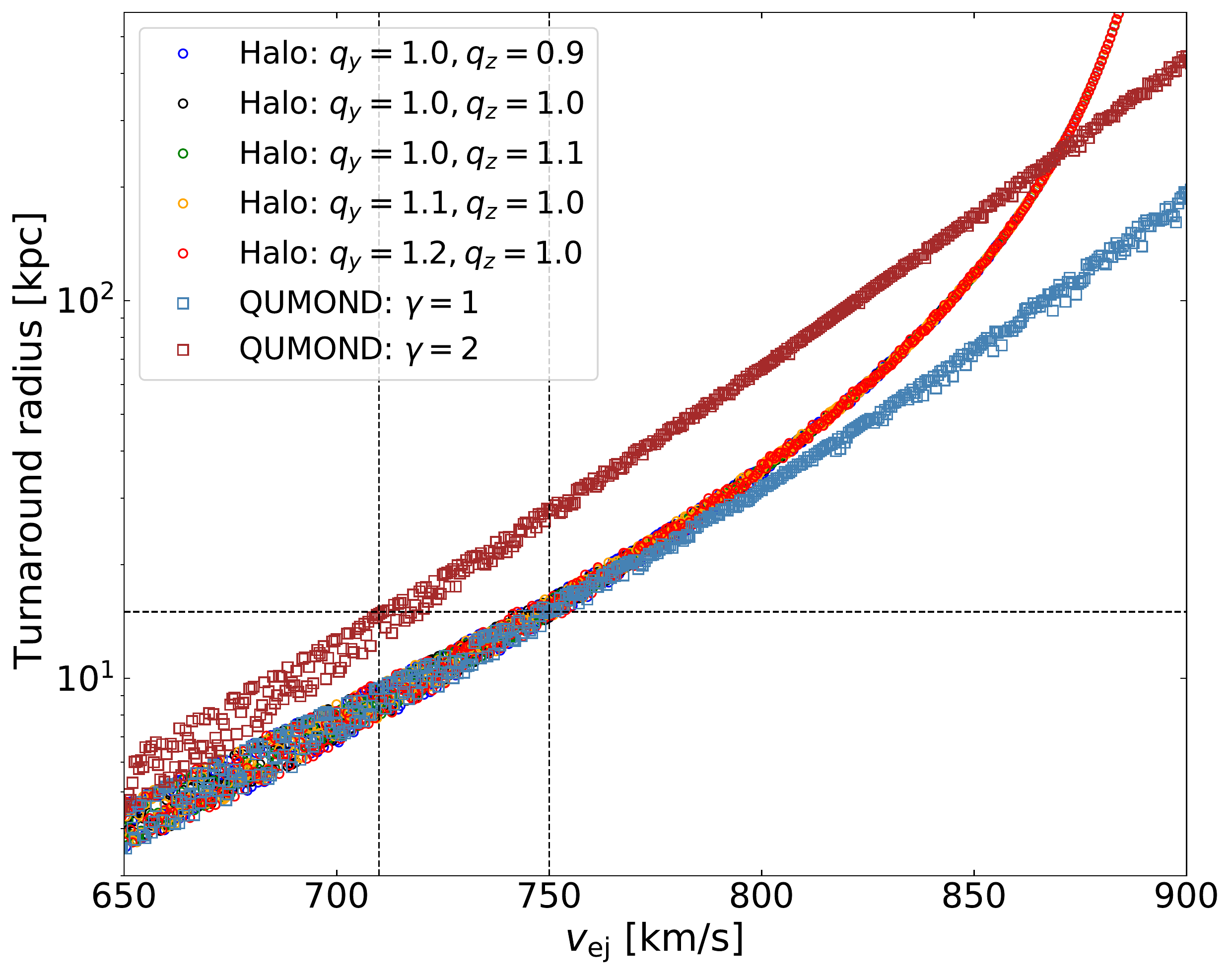}
\caption{\label{fig:vejrmax} HVS outer turnaround as a function of the ejection speed, $v_{\rm ej}$, in Newtonian gravity with different shapes of the dark matter halo and in QUMOND with $\gamma = 1$ and $\gamma=2$; models and symbols are detailed in the inset. The scatter of the points originates from the random initial directions of the HVS ejection velocities. The Newtonian gravitational potential is dominated by the dark matter halo that becomes approximately spherical at large radii and makes the scatter of the points decrease with increasing $v_{\rm ej}$. The vertical dotted lines mark the thresholds $v_{\rm ej}=710$ km~s$^{-1}$ for QUMOND with $\gamma = 2$ and $v_{\rm ej}=750$ km~s$^{-1}$ for all the other models;  the horizontal line marks the minimum galactocentric distance of 15~kpc, appropriate for 4~$M_{\sun}$ HVSs. Less massive stars have larger velocity thresholds and larger minimum galactocentric distances. }
\end{figure}

The ejection velocity is not observable but it is correlated with the outer turnaround, namely the maximum distance from the MW center that the star can travel. Figure~\ref{fig:vejrmax} shows the outer turnaround as a function of the ejection velocity of a star in  different models of the Galactic potential. This figure shows that neither in the Newtonian models nor in QUMOND with $\gamma = 1$ can stars with $v_{\rm ej} \lesssim 750$ km~s$^{-1}$ travel beyond a distance of about 15 kpc from the Galactic center.  Similarly, in QUMOND with $\gamma = 2$, the stars with $v_{\rm ej} \lesssim 710$ km~s$^{-1}$ cannot travel beyond 15 kpc. Therefore, identifying the $4~M_{\sun}$ stars that  have not experienced the inner turnaround, and thus have tangential velocities proportionate to the  deviation of the gravitational potential from axial symmetry, requires considering only $4~M_\sun$ HVSs at galactocentric distance $r > 15$ kpc.

The threshold ejection velocity increases with decreasing HVS mass, because of the longer lifetimes of lower-mass HVSs: longer-lived stars do not experience the inner turnaround only if they travel to larger galactocentric distances; in other words, only if they have larger ejection speeds.
For example, for $3~M_\sun$ or $2.5~M_\sun$ stars with lifetimes on the main sequence $\sim 350$ Myr and $\sim 580$ Myr, respectively \citep{schaller1992}, the threshold ejection velocities are  $\sim 790$ km~s$^{-1}$  and $\sim 815$ km~s$^{-1}$ in Newtonian models; these threshold ejection speeds  correspond to  outer turnaround radii of $\sim 30$ kpc  and $\sim 50$ kpc, respectively (Fig.~\ref{fig:vejrmax}). In QUMOND with $\gamma = 1$ ($\gamma = 2$), for $3~M_\sun$ or $2.5~M_\sun$ stars, the threshold ejection velocity is $\sim 800$ ($755$) km~s$^{-1}$ and $\sim 830$ ($785$) km~s$^{-1}$, respectively; the corresponding turnaround radius is $\sim 30$ ($30$) kpc and $\sim 50$ ($50$) kpc, respectively. Therefore, the $3~M_\sun$ or $2.5~M_\sun$ stars are of interest for our test only if they are at galactocentric distance $r$ larger than $30$ kpc and $50$ kpc, respectively.

\section{Tangential velocity in QUMOND and Newtonian gravity} 
\label{sec:angvel}

In this section we show the distributions of the azimuthal and latitudinal components of the tangential velocity, $v_\phi$ and $v_\theta$ respectively, in both QUMOND and Newtonian gravity. We show that, for HVSs within $\sim 60$ kpc,  QUMOND yields upper limits for $|v_\phi|$ values that are substantially lower than the values that non-axisymmetric gravitational potentials  can generate in Newtonian gravity. The QUMOND scenario might thus be challenged if values of $|v_\phi|$ higher than those upper limits are observed.

In QUMOND, the symmetry of the Galactic potential is determined by the distribution of the baryonic components alone, unlike the case of Newtonian gravity, where the shape of the dark matter halo plays a crucial and dominant role. In QUMOND, if the baryonic distribution is axially symmetric, the pull of the stellar disk generates non-null $v_\theta$ values, whereas $v_\phi$ is always zero. Non-null values of $v_\phi$ only appear in a non-axisymmetric distribution of baryons, which mostly originates from a triaxial bulge (Eq.~\ref{eq:bulge_tri}) or from a nonspherical HG halo (Sect.~\ref{sec:HG}). 

In Newtonian gravity with a perfectly axisymmetric distribution of the baryonic matter, as we adopt here (Eqs.~\ref{eq:pot_disk}-\ref{eq:pot_BH}), non-null values of $v_\phi$ can still appear if the halo of dark matter is not axially symmetric about the same axis of symmetry of the baryonic distribution, which is the $z$ axis in our models. More importantly, a tilted or triaxial dark matter halo acts on the variation in $v_\phi$ for most of the HVS trajectory, whereas in QUMOND the action of the triaxiality of the bulge is limited to the initial phase of the HVS trajectory within $\sim5$ kpc of the center. 

This fundamental difference suggests that we might expect substantially higher values of $v_\phi$ in Newtonian gravity than in QUMOND, unless in Newtonian gravity either the dark matter halo is perfectly spherical or the halo is axisymmetric and its axis of symmetry is aligned with the axis of symmetry of the baryonic components. These possibilities, however, appear unlikely in the current dark matter scenarios \citep{2007MNRAS.377...50H, 2013MNRAS.434.2971D}. 

A detailed study on how the shape of the dark matter halo in Newtonian gravity affects the velocity components of the HVSs is presented elsewhere \citep{Gallo}. Here, we briefly discuss the distributions of the azimuthal and latitudinal velocity components $v_\phi$ and $v_\theta$ in QUMOND and Newtonian gravity, and how $v_\phi$ can discriminate between the two theories of gravity.

\subsection{The upper limit of the azimuthal component, $v_\phi$, in QUMOND}
\label{sec:MOND}

In QUMOND, if the baryonic components are non-axisymmetric, the HVSs may have non-null $v_\phi$ values. To estimate these values of $\vert v_\phi\vert$, in Sect. \ref{sec:MONDBulge} we consider the effects of a triaxial bulge, which is the Galactic baryonic component that is expected to play the major role in affecting $|v_\phi|$ \citep{Gardner:2020jsf}. In Sect. \ref{sec:HG}, we illustrate the effects of including a nonspherical halo of hot gas surrounding a MW with a spherical bulge.

\subsubsection{The role of a triaxial bulge}
\label{sec:MONDBulge}

Finding the gravitational field of the triaxial bulge
over the entire numerical domain requires a demanding amount of computational time. We thus 
adopted a simplified approach that uses the fact that  the triaxiality of the bulge is not effective at distances $r \gtrsim$ 5 kpc, according to the density profile in Eq.~(\ref{eq:bulge_tri}) and the acceleration profile shown in Fig.~\ref{fig:bulgeacc}.  For our purpose, at these large radii, the gravitational potential of the bulge is well approximated by the potential generated by a spherically symmetric source. Therefore, we only 
used the field of a triaxial bulge to simulate the motion of the HVSs within 5 kpc of the Galactic center; at larger radii, we adopted the gravitational field of a spherical bulge. We matched the two regimes by setting the components of the velocity at $r=5$ kpc from the inner region as the initial conditions for the outer region. The gravitational acceleration at $r=5$~kpc due to the triaxial bulge is about 7\% smaller than  the gravitational acceleration at $r=5$~kpc due to the spherical bulge, suggesting that our approach introduces a limited error in the integration of the equation of motion of the HVSs. In fact, this approximation does not affect our main result: it produces a slight overestimate of the QUMOND upper limit we discuss below, because the acceleration of the spherical bulge is higher than the actual triaxal bulge acceleration.

As argued in Sect.~\ref{sec:sample}, $4~M_{\sun}$ stars with ejection velocity $v_{\rm ej}\lesssim 710$~km~s$^{-1}$ in QUMOND with $\gamma = 2$, or with $v_{\rm ej}\lesssim 750$~km~s$^{-1}$ in the other QUMOND and Newtonian models, tend to have disproportionately high tangential velocity components and cannot be used to probe the nonspherical components in the Galaxy potential. The maximum value of $v_\phi$ is inversely proportional to the ejection velocity, as expected by the argument illustrated in Sect.~\ref{sec:sample}. 

\begin{figure}[!ht]
\centering 
\includegraphics[width=8cm]{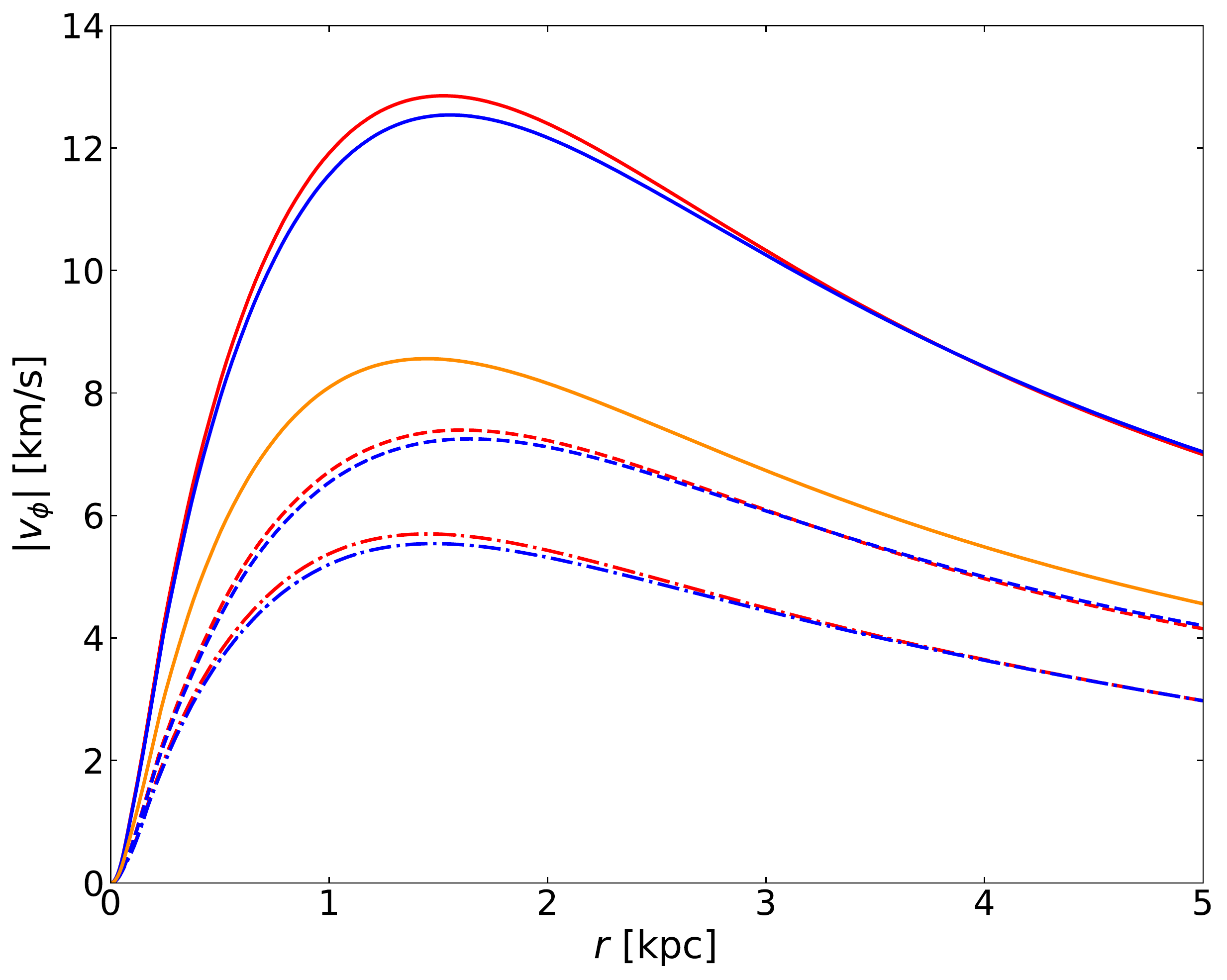}
\caption{\label{fig:bulgevphi} Magnitude of the azimuthal velocity component, $v_\phi$, in QUMOND as a function of the galactocentric distance, $r$, for three 
4~$M_{\sun}$ HVSs with different ejection velocities: $v_{\rm ej} = 710$ km~s$^{-1}$ in QUMOND with $\gamma = 2$ (red), $v_{\rm ej} = 750$ km~s$^{-1}$ in QUMOND with $\gamma = 1$ (blue), and $v_{\rm ej} =1000$ km~s$^{-1}$ in QUMOND with $\gamma = 1$ (orange). The solid lines are for the stars ejected along $\theta = 45^\circ$ and $\phi = 45^\circ$, whereas the dashed (dash-dotted) line is for $\theta = 45^\circ$ and $\phi = 15^\circ$ ($75^\circ$). In these examples, the QUMOND field is due to the triaxial bulge and the axisymmetric disk. The stars acquire $v_\phi $ only due to the triaxiality of the bulge. }
\end{figure}

In order to find the maximum $\vert v_\phi \vert$ in QUMOND, we chose the lowest ejection velocity relevant for our analysis. Figure~\ref{fig:bulgevphi} shows $\vert v_\phi \vert$ as a function of the galactocentric distance $r$ for three different ejection velocities: $v_{\rm ej} = 710$ km~s$^{-1}$ in QUMOND with $\gamma = 2$ (red curves), $v_{\rm ej} = 750$ km~s$^{-1}$ and 1000 km~s$^{-1}$ in QUMOND with $\gamma = 1$ (red and orange curves). All the stars shown in Fig.~\ref{fig:bulgevphi} are ejected at a polar angle $\theta=45^\circ$; the solid lines show $v_\phi$ for an azimuthal angle $\phi=45^\circ$, whereas the dashed and dot-dashed lines show $v_\phi$ for the slowest stars for different $\phi$. We remind the reader that the largest semimajor axis of the triaxial bulge is along the $x$ axis (see Eq.~\ref{eq:bulge_tri} and Fig.~\ref{fig:vcomp}). According to Fig.~\ref{fig:bulgevphi}, at $r=5$ kpc, the maximum $v_\phi$ we should expect is $\sim 8$ km~s$^{-1}$. We note that this result is not sensitive to the choice of the interpolating function $\nu$ (Eq.~\ref{eq:nu}), because within 5 kpc the gravitational field is mostly Newtonian.

\begin{figure}
\centering 
\includegraphics[width=8cm]{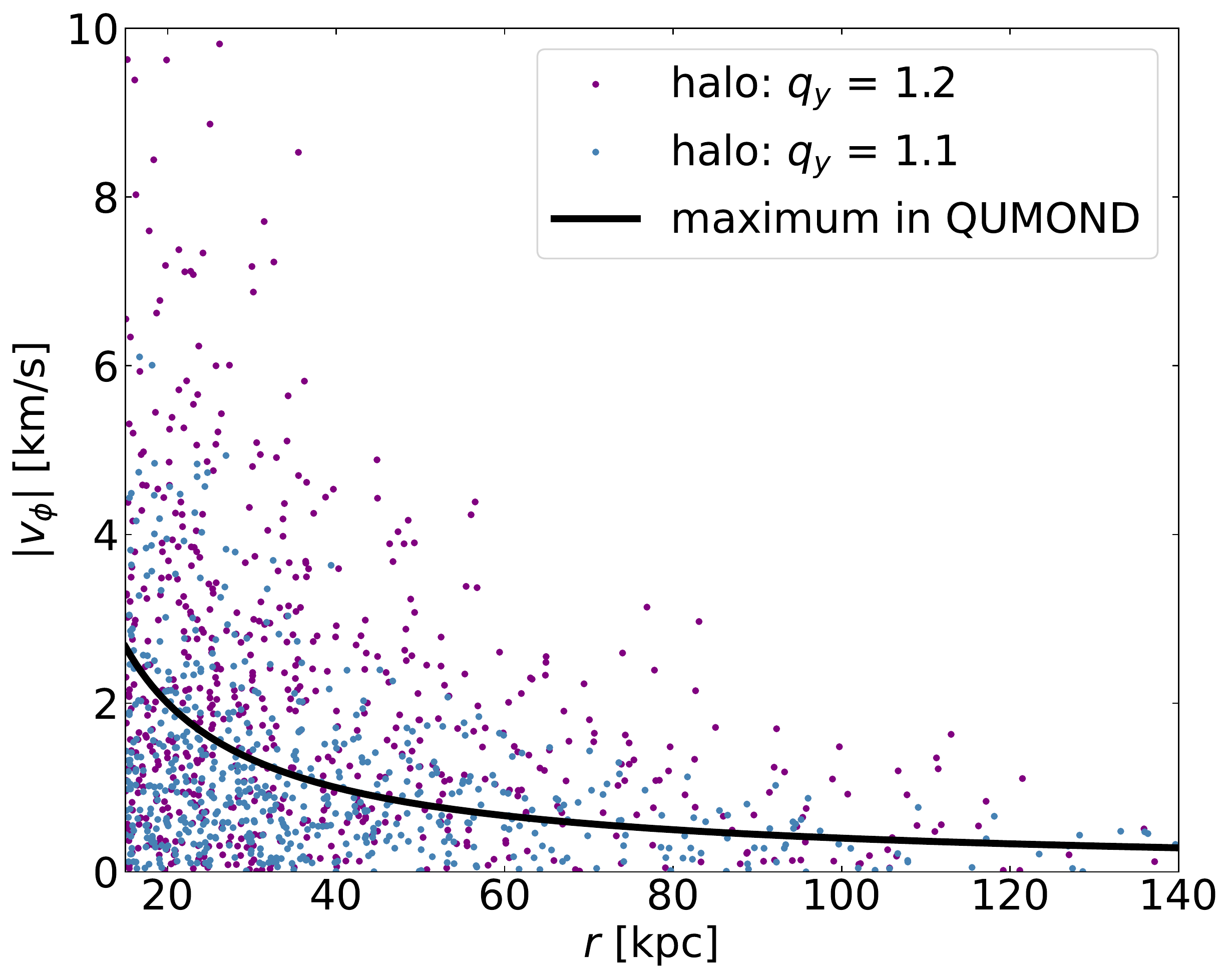}
\caption{\label{fig:rvphi} Magnitudes of the azimuthal velocity components $v_\phi$ of $4~M_{\sun}$ HVSs in Newtonian gravity as a function of their radial coordinates, $r$, at the time of observation for a prolate dark matter halo with axis of symmetry in the plane of the disk. The light blue and purple dots show the HVSs for a halo with $(q_y, q_z) = (1.1, 1)$ and $(q_y, q_z) = (1.2, 1)$, respectively. The black line shows the upper limit of $|v_\phi|$ in QUMOND due to the triaxial bulge. The upper limit also holds for HVSs with masses lower than 4~$M_{\sun}$,  provided they are beyond a certain galactocentric distance, e.g., $r \gtrsim 30$ kpc (50 kpc) for $3~M_\sun$ ($2.5~M_\sun$) HVSs.}
\end{figure}

\begin{figure*}[!h]
\centering
\includegraphics[width=8.5cm]{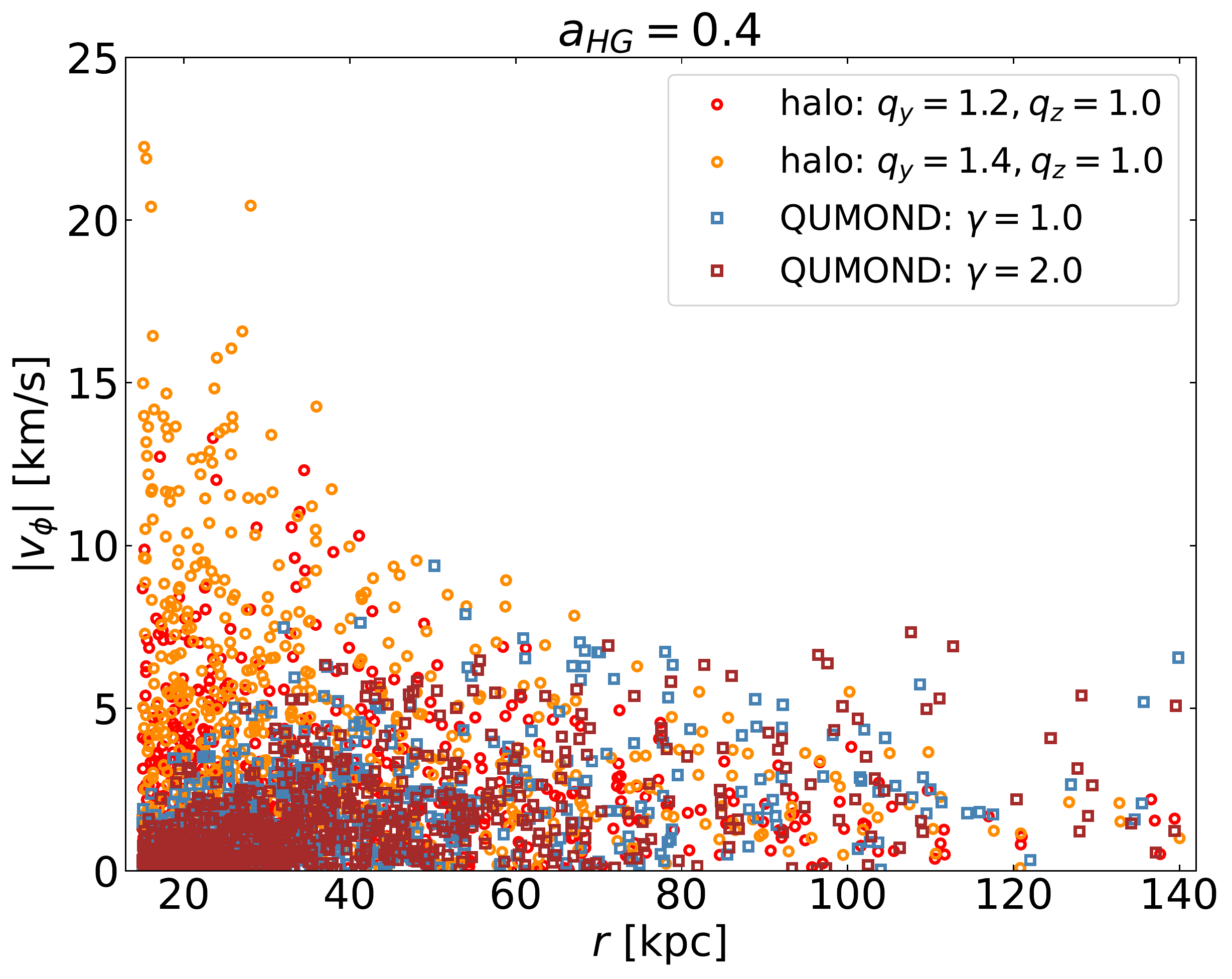}
\includegraphics[width=8.5cm]{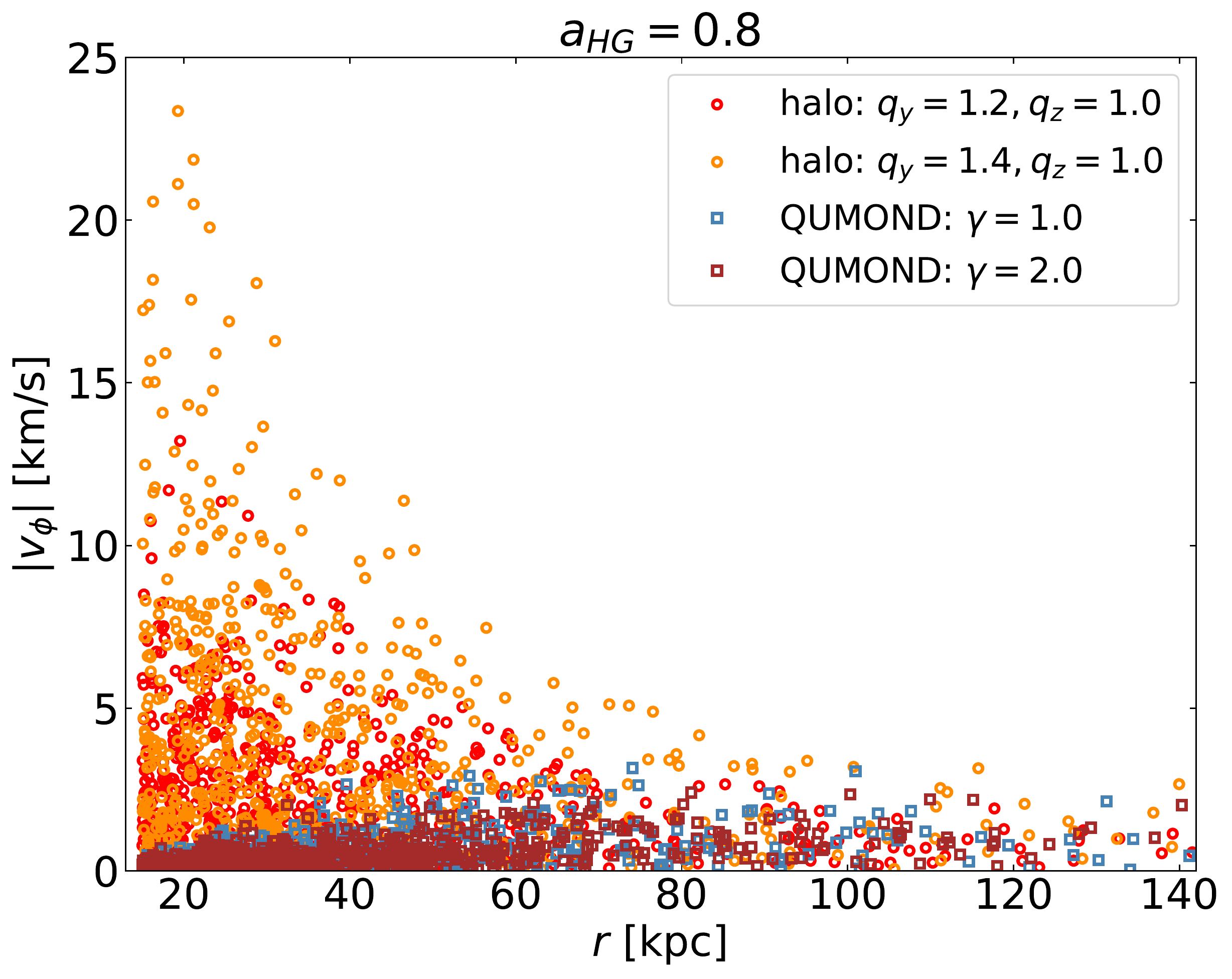}
\caption{Magnitudes of the azimuthal velocity components, $v_\phi$, of $4~M_{\sun}$ HVSs as a function of their radial coordinates, $r$, at the time of observation in the presence of an HG halo with total mass of $1.5 \times 10^{11} M_\odot$ within 100 kpc in both Newtonian and QUMOND scenarios. The left and right panels show the results for two different oblatenesses, $a_{\rm HG} =$ 0.4 and 0.8, of the HG halo (see Eqs.~\ref{eq:rhoHG}-\ref{eq:mHG}).  }
\label{fig:rvphiHG10}
\end{figure*}

\begin{figure*}[!h]
\centering
\includegraphics[width=8.5cm]{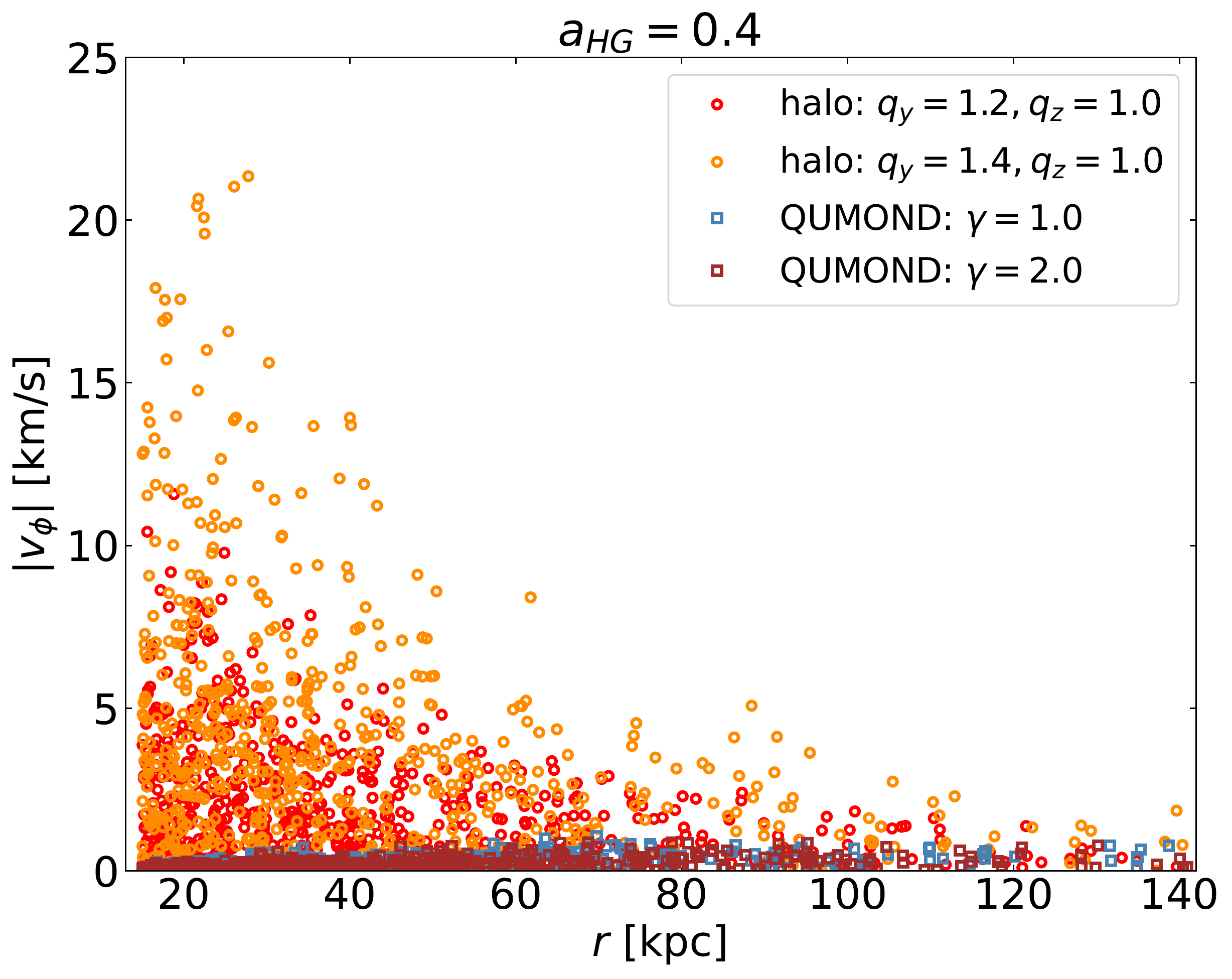}
\includegraphics[width=8.5cm]{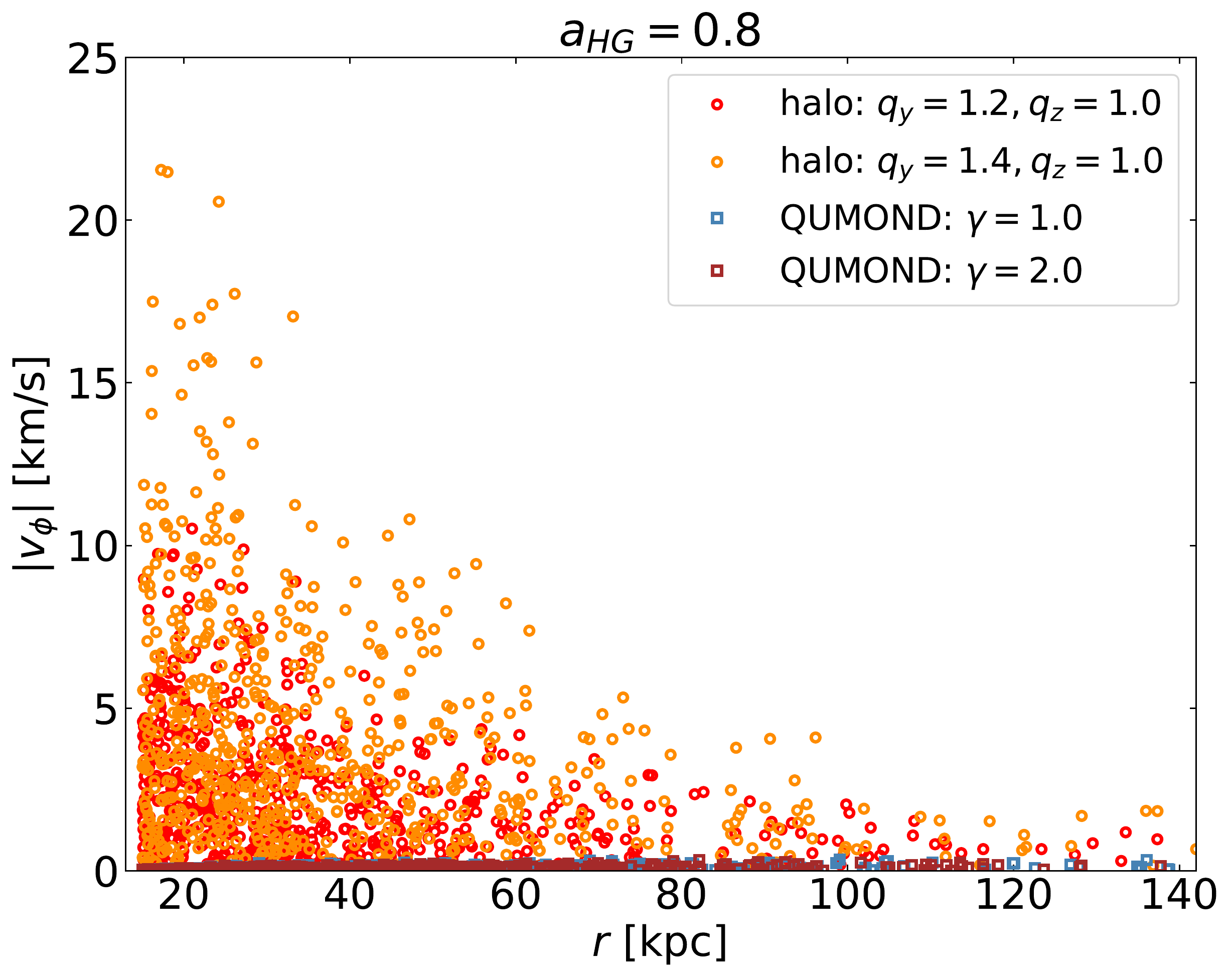}
\caption{Same as Fig.~\ref{fig:rvphiHG10} except for the HG halo with a lower mass: $1.5 \times 10^{10} M_\odot$ within 100 kpc.}
\label{fig:rvphiHG01}
\end{figure*}

\begin{figure*}
\centering 
\includegraphics[width=14cm]{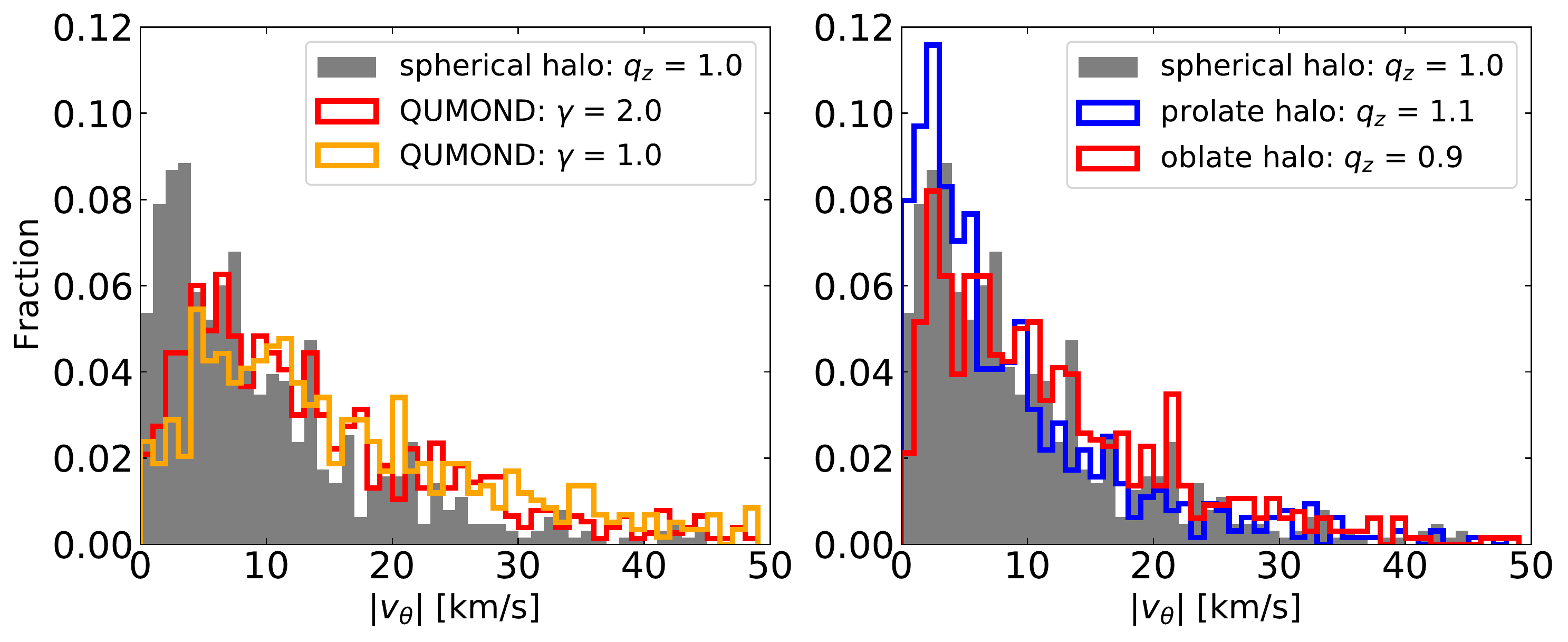}
\caption{\label{fig:histvth} Distributions of the latitudinal component of the velocity, $\vert v_\theta\vert$, in Newtonian gravity and in QUMOND for 4~$M_{\sun}$ HVSs.
The left panel shows the distributions of $\vert v_\theta\vert$  for different models of the Galactic potential: Newtonian gravity with a spherical dark matter halo (gray shaded histogram) and QUMOND with $\gamma = 1$ (orange histogram) and $\gamma = 2$ (red histogram). As the gravitational pull of the baryonic disk is enhanced in QUMOND, the fraction of HVSs with high $\vert v_\theta\vert$ is larger than in Newtonian gravity with a spherical halo. The right panel shows the distributions of  $\vert v_\theta\vert$  for three different shapes of the dark matter halo in Newtonian gravity. The three halos have $q_y = 1$ but different $q_z$: a spherical halo with $q_z = 1$ (gray shaded histogram), a prolate halo with $q_z = 1.1$ (blue histogram), and an oblate halo with $q_z = 0.9$ (red histogram). An oblate  halo enhances the gravitational pull of the baryonic disk, as shown by the  larger fraction of HVSs with high $\vert v_\theta\vert$ and the smaller fraction of HVSs with low $\vert v_\theta\vert$ (compare the shaded and red histograms). The opposite occurs with a prolate halo (compare the shaded and blue histograms).}
\end{figure*}

Assuming that the angular momentum per unit mass $\vert\ell_z\vert = r \sin \theta \vert v_\phi\vert$ is conserved at larger radii, and $\vert v_\phi\vert$ thus falls off as $r^{-1}$, we obtain
\begin{equation}
\vert v_\phi ^{\rm max}(r)\vert \sim 4 \ \left( \frac{10\ {\rm kpc}} {r} \right) {\rm km\ s}^{-1}\, 
\label{eq:QUMONDupperlimit}
\end{equation}
for 4 $M_{\sun}$ stars. The upper-limit radial profile of $\vert v_{\phi} \vert$ reported in Eq.~(\ref{eq:QUMONDupperlimit}) is shown by the black solid line in Fig.~\ref{fig:rvphi}. The above equation provides a conservative upper limit for two reasons.
First, the equation was derived using the 4 $M_{\sun}$ HVSs with the lowest ejection velocities relevant for our analysis, namely 710 and 750 km~s$^{-1}$. Any HVS that travels beyond 15 kpc (see Fig.~\ref{fig:vejrmax}) has higher ejection velocity, hence lower azimuthal velocity. Secondly, the star trajectories bend toward the Galactic disk located in the $x$-$y$ plane ($\theta = 90^\circ$; see Fig.~\ref{fig:vcomp}), so that $\theta$ approaches $90^\circ$.\footnote{As the disk is located on $\theta = 90^\circ$, $\theta$ always approaches $90^\circ$. For an HVS ejected above the disk, the initial value of $\theta$ is between $0^\circ$ and $90^\circ$, and $\theta$ increases. When the HVS is ejected below the disk, the initial $\theta$ is between $90^\circ$ and $180^\circ$, and $\theta$ decreases.} 
As a result, $\sin \theta$ always increases and causes a further decrease in $\vert v_\phi\vert$, because $\vert\ell_z\vert = r \sin \theta \vert v_\phi\vert$ remains constant. Therefore, using the HVS with the lowest ejection velocity and taking $\sin\theta$ as a constant yield the conservative upper limit of $\vert  v_\phi\vert$ given in Eq.~(\ref{eq:QUMONDupperlimit}). 

The radial profile $\vert v^{\rm max}_{\phi} (r)\vert$  of Eq.~(\ref{eq:QUMONDupperlimit}) also provides a conservative upper limit  for stars with masses lower than 4~$M_{\sun}$. Indeed, as discussed in Sect.~\ref{sec:sample}, stars with lower mass, and thus longer lifetimes, must  be ejected with higher speeds in order to avoid the inner turnaround. Therefore, for higher ejection speeds, the normalization of the azimuthal velocity radial profile, $\vert v_{\phi} (r)\vert$ in Eq.~(\ref{eq:QUMONDupperlimit}), will be lower.

In Newtonian gravity with an axisymmetric distribution of baryons (Eqs.~\ref{eq:pot_disk}-\ref{eq:pot_BH}), the HVSs obtain non-null $\vert v_\phi\vert$ values only when the dark matter halo is non-axisymmetric  about the $z$ axis, as in our models with $q_y \neq 1$. Figure~\ref{fig:rvphi} shows the values of $\vert v_\phi \vert$ and $r$ of the HVSs for two different shapes of the halo: $(q_y, q_z) = (1.1, 1)$, and $(q_y, q_z) = (1.2, 1)$. In both cases,  the halo is prolate with the semimajor axis along the $y$ axis in the plane of the disk. The halo becomes more spherical at larger radii, $r \gtrsim r_a = 21.6$ kpc. Thus, the azimuthal angular momentum, $\ell_z = r \sin \theta \ v_\phi$, of each HVS becomes a constant as the star travels beyond $r \sim r_a = 21.6$ kpc  and $\vert v_\phi\vert $ decreases as $r^{-1}$. Near $r \sim 21.6$ kpc, $\vert v_\phi \vert$ of an HVS can be as high as 10 km~s$^{-1}$ for $q_y = 1.2$, and 6 km~s$^{-1}$ for $q_y = 1.1$ (see Fig.~\ref{fig:rvphi}). These values are substantially higher than the QUMOND upper limit $v_\phi^{\rm max}(r)\sim 1.8$ km~s$^{-1}$  derived with Eq.~(\ref{eq:QUMONDupperlimit}) at that distance and shown by the solid line in Fig.~\ref{fig:rvphi}.

\subsubsection{The effect of a hot gaseous halo}
\label{sec:HG}

In addition to the baryonic components considered in the previous section, we included in our MW model a reservoir of baryons in the form of an HG halo extending up to the virial radius of the MW. The existence of this HG halo with a temperature $\sim 10^6$ K is suggested by the O VII and O VIII emission and absorption lines in the soft X-ray band \citep{2003ARA&A..41..291P, 2012ApJ...756L...8G, 2013ApJ...762...20F, 2013MNRAS.433.2749G, 2015ApJ...815...77S}. The presence of such diffuse hot gas may alleviate the missing baryon problem \footnote{The baryonic census in the Local Group appears to add up to only $\sim 15\%$ \citep{1998ApJ...503..518F} of the baryonic mass expected from the estimated baryonic abundance ($\Omega_b h^2 \approx 0.022$) \citep{2020A&A...641A...6P} from the Big Bang nucleosynthesis and the CMB anisotropies. } \citep{1998ApJ...503..518F, 2012ApJ...756L...8G, 2012ApJ...759...23S, 2013ApJ...762...20F, 2020A&A...641A...6P}. An oblate gaseous halo with the smallest principal axis lying on the Galactic plane may be  part of the Vast Polar Structure of the MW \citep{2011A&A...532A.118P, 2013A&A...557L...3Z, 2013MNRAS.431.3543H}.

We modeled the HG halo using the density profile \citep{2017A&A...603A..65T}
\begin{equation}
\rho_{\rm HG} (m) = \rho_{\rm 0,HG} \left(1+\frac{m}{r_{\rm 0,HG}}\right)^{-3} \exp\left(- \frac{m^2}{r_{\rm t,HG}^{2}}\right)\, , \label{eq:rhoHG}
\end{equation}
where $r_{\rm 0,HG} = 100$ kpc is the core radius and $r_{\rm t,HG} = 200$ kpc is the truncation radius. The elliptical radius $m$ for the oblate halo is
\begin{equation}
m = \sqrt{\frac{x^2}{a_{\rm HG}^2} + y^2 + z^2} \ . \label{eq:mHG}
\end{equation}
Currently, the shape of the halo has no observational constraints; we thus varied $a_{\rm HG}$ from 0.4 to 0.8. Similarly, its total mass is uncertain by a factor of ten \citep{2013ApJ...770..118M, 2013MNRAS.433.2749G, 2015ApJ...815...77S}, and we thus explored two different values of $\rho_{\rm 0,HG}$: $3.0 \times 10^5$ and $3.0 \times 10^4$ $M_\odot$ kpc$^{-3}$. These values yield a total mass of $1.5 \times 10^{11} M_\odot$ and $1.5 \times 10^{10} M_\odot$, respectively, within 100 kpc.

We explored the effects of the HG halo on the azimuthal velocities of the HVSs in both the Newtonian and QUMOND scenarios (Figs.~\ref{fig:rvphiHG10}-\ref{fig:rvphiHG01}). In both cases, we considered the axisymmetric models for the central black hole, bulge and disk of Eqs.~(\ref{eq:pot_disk}-\ref{eq:pot_BH}). 

In Newtonian gravity, we considered the non-axisymmetric dark matter halos with $(q_y , q_z) = (1.2, 1)$ and $(q_y, q_z)=(1.4, 1)$ and assumed that the principal axes of the dark halo coincided with that of the HG halo. Whereas the dark halo is nearly spherical at large radii ($r \gtrsim 21.6$ kpc) and its contribution to the $|v_{\phi}|$ values thus falls as $r^{-1}$ at large distances, the constant oblateness of the HG halo increases the $|v_{\phi}|$ values at all radii. Nevertheless the $|v_{\phi}|$ values are dominated by the shape of the non-axisymmetric dark matter halo, because its mass is at least 10 times the mass of the HG halo. 

On the contrary, in QUMOND, the gravitational field is substantially enhanced at large distances by the presence of the HG halo and its mass and shape have a relevant effect on the values of $|v_{\phi}|$. Therefore, for the HG halo with the highest mass $M(<100\, \mathrm{kpc}) = 1.5 \times 10^{11}\, M_\odot$ (Fig.~\ref{fig:rvphiHG10}), although the $|v_\phi|$'s in Newtonian scenario can still exceed the values in QUMOND at distances between 15 kpc and 60 kpc irrespective of the shape of the HG halo, at larger distances, the maximum possible values of $|v_\phi|$ in QUMOND are higher than or comparable to the maximum possible values in Newtonian gravity, depending on the shape of the HG halo. On the contrary, for the HG halo with the lowest mass $M(<100\, \mathrm{kpc}) = 1.5 \times 10^{10}\, M_\odot$ (Fig.~\ref{fig:rvphiHG01}), the $|v_\phi|$ values in Newtonian gravity can be significantly higher than the values in QUMOND at distances between 15 kpc and 100 kpc.

In summary, the presence of an oblate HG halo increases the upper limit of the $|v_\phi|$'s in QUMOND at all distances, compared to the presence of a triaxial bulge alone. However, at smaller distances (15 kpc $\lesssim r \lesssim$ 60 kpc), the $|v_\phi|$ values in Newtonian gravity due to a non-axisymmetric dark matter halo may still substantially exceed the QUMOND values.

\subsection{Latitudinal component, $v_\theta$}
\label{sec:vthMOND}

The difference in the latitudinal component $v_\theta$ of the velocity in QUMOND and in Newtonian gravity is subtler than the difference in $v_\phi$. In QUMOND, the source of the gravitational field is concentrated in the plane of the disk and it bends the HVSs trajectories toward this plane. In Newtonian gravity, the main role is played by the spheroidal dark matter halo: A prolate halo, with its major axis perpendicular to the stellar disk, is likely to generate $v_\theta$  lower than in QUMOND; on the contrary, an oblate halo with its major axis in the plane of the disk will generate $v_\theta$ comparable or even higher than in QUMOND. Accurate predictions of these differences clearly depend on the exact values of the axial ratios, in addition to the actual mass and size of the dark matter halo.

From our knowledge of the baryonic components, we can predict the distribution of $\vert v_\theta\vert$ in Newtonian gravity, assuming that the dark matter halo is spherical and therefore it does not affect $v_\theta$. Similarly, in our adopted model, the SMBH and the bulge have spherically symmetric potential and do not contribute to $v_\theta$. The shaded histogram in the left panel of Fig.~\ref{fig:histvth} shows the distribution of $ \vert v_\theta\vert $ in this Newtonian model, effectively caused by the disk alone. In QUMOND, at larger distances, where the Newtonian gravitational field approaches $a_0$, the gravitational pull of the disk gets enhanced compared to Newtonian gravity. Consequently, the HVSs have larger $\vert v_\theta\vert$ in QUMOND than in Newtonian gravity.

\begin{figure}
\centering 
\includegraphics[width=8cm]{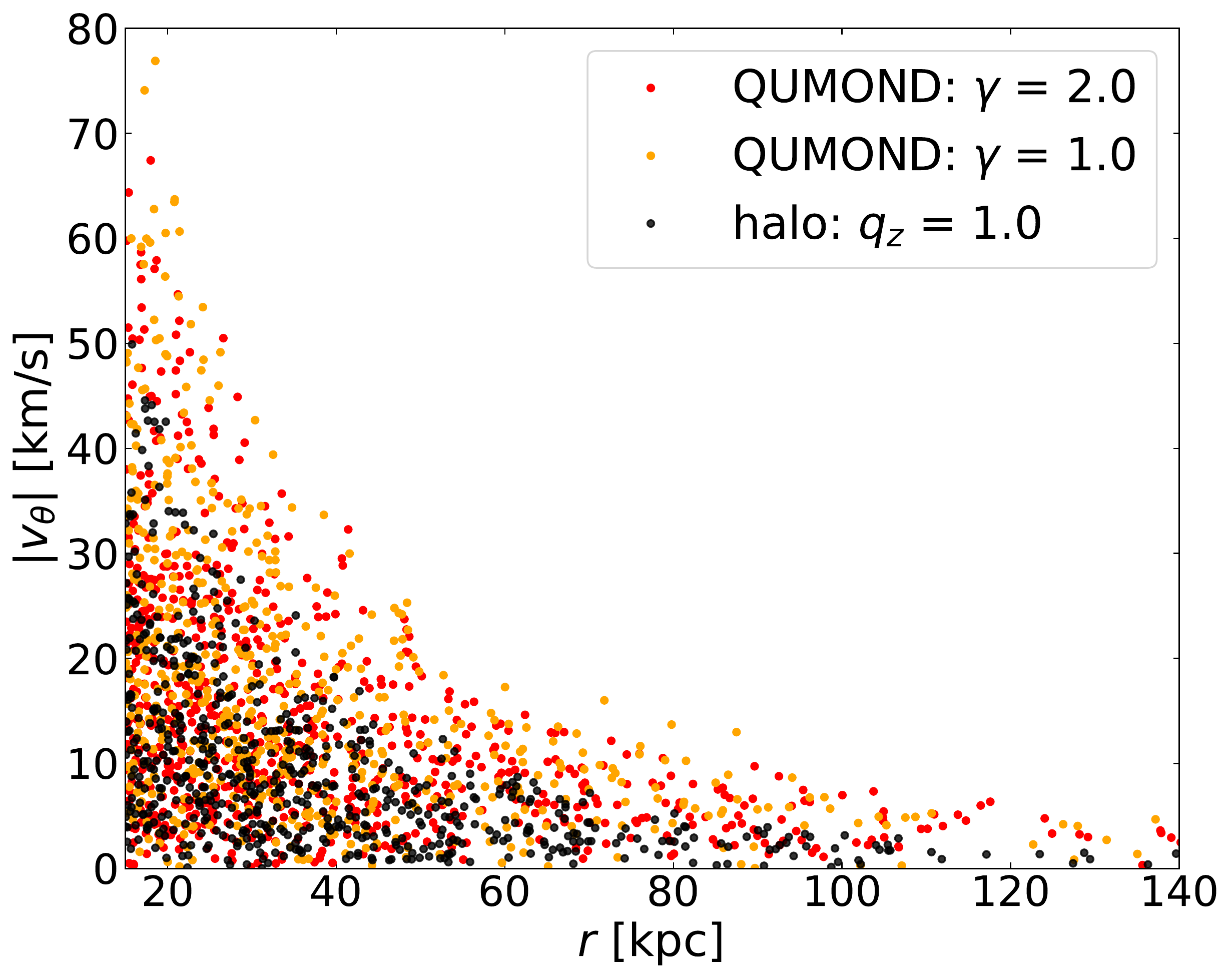}
\caption{\label{fig:rvthMOND} Latitudinal velocity components, $|v_\theta|$, as a function of the galactocentric distance, $r$, of 4~$M_{\sun}$ HVSs in QUMOND with $\gamma = 1$ (orange dots) and $\gamma=2$ (red dots), and in Newtonian gravity with a spherical dark matter halo ($q_z = 1$) (black dots).}
\end{figure}

Figure~\ref{fig:rvthMOND} shows the magnitude of $v_\theta$ as a function of the galactocentric distance $r$ due to the baryonic components in QUMOND with $\gamma = 1$ and 2 (orange and red dots, respectively), and in Newtonian gravity with a spherical halo (black dots). In both models, the maximum possible values of $\vert v_\theta\vert$ decrease as $\sim r^{-1}$, because, as each star travels beyond the length scale of the stellar disk, its angular momentum, $\sim r v_\theta$, becomes a constant. Because of the enhanced pull of QUMOND, the maximum value of $v_\theta$ is higher than in Newtonian gravity at all radii. Unfortunately, unlike the case of $v_\phi$, these differences cannot be used to distinguish QUMOND from Newtonian gravity, because these same differences can be generated  by an appropriate  shape of the dark matter halo, as we illustrate below.

In Newtonian gravity with a dark matter halo, non-null $v_\theta$ values are caused by the Galactic disk and by the halo with $q_z \neq 1$ (models sketched in panels (c) and (d) of Fig.~\ref{fig:MW}). For an oblate halo ($q_z < 1$), the gravitational pull of the halo enhances the pull of the disk. Hence, the number of HVSs with high $\vert v_\theta\vert$ is larger than in the case of a spherical halo, as shown by the comparison of the red and shaded histograms in the right panel of Fig.~\ref{fig:histvth}. The reverse happens for a prolate halo with $q_z > 1$. Therefore, the difference between the QUMOND and the Newtonian distributions shown in the left panel of Fig.~\ref{fig:histvth} can be easily mimicked in Newtonian gravity by an appropriate oblate dark matter halo.

\begin{figure}
\centering 
\includegraphics[width=8cm]{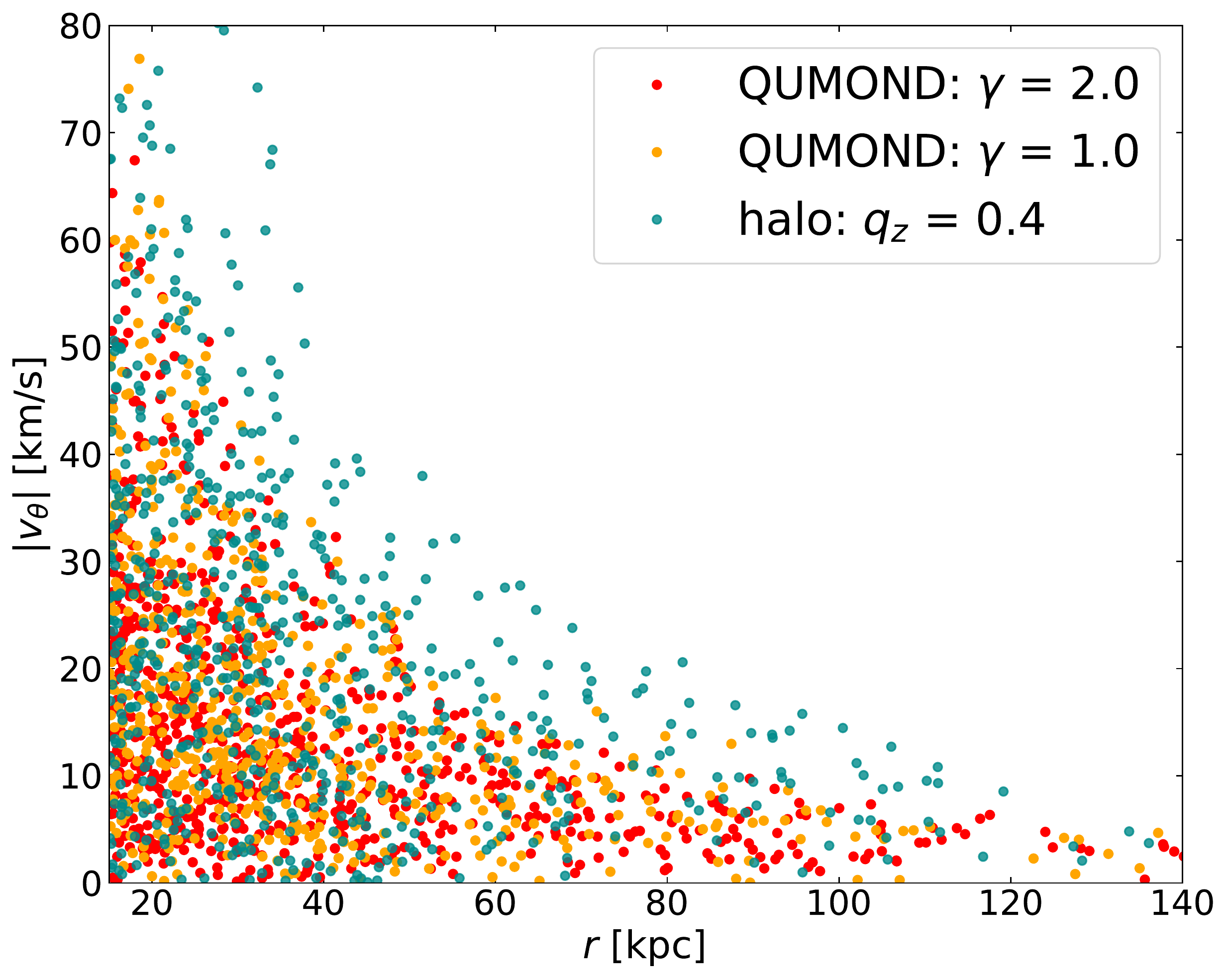}
\caption{\label{fig:rvthMONDOblate} Latitudinal velocity components, $|v_\theta|$, as a function of the galactocentric distance, $r$, of 4~$M_{\sun}$ HVSs in QUMOND with $\gamma = 1$ (orange dots) and $\gamma=2$ (red dots), and in Newtonian gravity with an oblate dark matter halo with $q_z = 0.4$ (dark cyan dots).}
\end{figure}

This degeneracy was emphasized by \citet{2005MNRAS.361..971R} when they attempted to distinguish MOND from Newtonian gravity using the stellar streams of the Sagittarius dwarf. The degeneracy between an oblate halo and QUMOND can be broken if $q_z$ is sufficiently small. Indeed, similar to the case of $v_\phi$, QUMOND sets an upper limit to $|v_\theta|$; on the contrary, in Newtonian gravity, $|v_\theta|$ may be higher than this upper limit if $q_z$ is sufficiently small. In our models, the Newtonian $|v_\theta|$ values are higher than the QUMOND upper limit if $q_z \lesssim 0.6$. However, the difference between the QUMOND and Newtonian $|v_\theta|$ values is not as prominent as for  $|v_\phi|$ values: In our model, the transition scale of the halo shape from oblate to spherical is $r_{\rm a} = 21.6$ kpc; therefore, the size of the portion of the halo that is actually oblate is comparable to the size of the baryonic disk. On the contrary, the size of the triaxial bulge causing nonzero $|v_\phi|$ values in QUMOND is much smaller than the scale length of the non-axisymmetric halo responsible for non-null $|v_\phi|$'s in Newtonian gravity. Therefore, $v_\theta$ is less effective than $v_\phi$ at distinguishing QUMOND from Newtonian gravity. Figure~\ref{fig:rvthMONDOblate} quantifies this difficulty by showing $|v_\theta|$ versus $r$ for QUMOND with $\gamma = 1$ and $\gamma=2$, and for Newtonian gravity with an oblate halo with $q_z = 0.4$; this dark matter halo is a rather extreme case, when compared to the current estimates of the shape of the MW dark matter halo \citep{2014ApJ...794..151L}. 

\subsection{Azimuthal velocities: A comparison with real data} 

We attempted a first comparison of the QUMOND upper limit $\vert v_\phi^{\rm max} (r)\vert$ (Eq.~\ref{eq:QUMONDupperlimit}) with the measured $\vert v_\phi^{\rm obs}\vert$ of nine stars drawn from the HVS survey sample of \cite{brown2014}. These nine stars have masses in the range $\sim 2.5-4~M_{\sun}$.
Their galactocentric distances are larger than the minimum mass-dependent radii required for our test (see Sect.~\ref{sec:sample}). In addition, \citet{kenyon2018} show that HVSs ejected within $25^\circ$ from the line joining the MW center and the LMC are affected by the LMC when they reach a galactocentric distance of $35$ to $65$ kpc. The nine HVSs of our sample  are all located more than $\sim 84^\circ$ away from the LMC, and the LMC pull should thus be irrelevant.

To be used for our test, these stars must originate from the Galactic center rather than being disk runaway stars. In principle, we could distinguish the two kinds of stars by tracing their trajectories back in time, if we knew the correct theory of gravity and the correct MW gravitational potential. When this information is unknown, and it is actually what we wish to constrain, this approach clearly generates a circularity problem. We  can solve this problem by tracing back the star trajectories in different gravitational potentials and different theories of gravity to select those stars, if any, that appear to be HVSs in all models. We will investigate this self-consistent treatment, and thus the actual feasibility of our test,  elsewhere. Here, we simply wished to see whether, if we assumed that the observed allegedly HVSs were indeed HVSs in all the models, the current data would have sufficed to distinguish between QUMOND and Newtonian gravity.

We relied on the analysis of \citet{brown2018} who, in Newtonian gravity, assume an axisymmetric disk and a spherical dark matter halo \citep{kenyon2014} to estimate a larger probability for the nine stars mentioned above to come from the Galactic center than to be disk  runaway stars.

We computed the azimuthal components, $v_\phi^{\rm obs}$, of their galactocentric velocities from the proper motions available in the Gaia Early Data Release 3 \citep[EDR3;][]{gaia2016b,gaiaEDR3}. We adopted the heliocentric distances derived from \citet{brown2018}, and the radial velocities of \cite{brown2014}. We 
found that  the magnitudes of the azimuthal velocities are in the range $\vert v_\phi^{\rm obs} \vert \simeq 37-452$~km~s$^{-1}$. For eight of these stars, the uncertainties on the azimuthal velocities are dominated by the errors on the proper motions, whereas the errors on the stars' distances and radial velocities, as well as on the position and velocity of the Sun in the galactocentric reference frame, appear to be negligible. For these stars, the relative uncertainties on the azimuthal velocities are in the range $\sim 50-340$\%. For the ninth star, B329, neglecting the error on its distance is inappropriate. B329 is a $(3.21 \pm 0.24)~M_{\sun}$ star at galactocentric distance $r=(61 \pm 13)$ kpc; by ignoring the uncertainty on the distance, we derived  
$\vert v_\phi^{\rm obs} \vert = (371 \pm 74)$~km~s$^{-1}$, whereas including the uncertainty on the distance yields $\vert v_\phi^{\rm obs} \vert = (371 \pm 131)$~km~s$^{-1}$. 

We conclude that the current uncertainties on $\vert v_\phi^{\rm obs} \vert$ make the azimuthal components consistent with zero within 3$\sigma$ for all of the nine stars of our sample. We are thus unable to verify whether these measures would be in principle consistent with the QUMOND limit. Indeed, even in Newtonian gravity, the large values of $\vert v_\phi^{\rm obs} \vert$ would probably imply an unrealistically large flatness or triaxiality of the dark matter halo (Fig.~\ref{fig:rvphi}). Therefore, even if these nine stars were HVSs in both QUMOND and Newtonian gravity, the comparison of their azimuthal components with the QUMOND upper limit, $\vert v_\phi^{\rm max} (r)\vert$, would still be inconclusive.

\section{Discussion and conclusions}
\label{sec:conclusions}

We showed that  measuring the galactocentric tangential velocity of HVSs in the MW can effectively allow us to  discriminate  between MOND, in its QUMOND formulation, and Newtonian gravity. Specifically, we 
demonstrated that HVSs with sufficiently high ejection speed possess galactocentric azimuthal velocities whose magnitude, $|v_\phi|$, cannot exceed a velocity threshold in QUMOND, while $|v_\phi|$ has no upper bounds in Newtonian gravity. This result could naturally translate into an observational test to discriminate between these two theories of gravity.

Our findings follow from the fact that the HVSs ejected from the Galaxy center on radial orbits acquire non-null azimuthal tangential speeds, $v_\phi$, due to the non-axisymmetric components of the Galactic gravitational potential. Hypervelocity stars with low ejection velocity can turn back toward the Galactic center and move outward again, acquiring a high $|v_\phi|$ that is not proportionate to the deviation from the axial symmetry of the potential. In contrast, HVSs with high ejection velocities reach larger galactocentric distances before turning back and thus die before experiencing the inner turnaround. These HVSs acquire a substantially lower $|v_\phi|$, which is proportional to the deviation from the axial symmetry of the potential.

In our models of the Galactic gravitational potential, the ejection velocity threshold for $4~M_{\sun}$ stars is $\sim 710$~km~s$^{-1}$ in QUMOND with $\gamma = 2$, and it is $\sim 750$~km~s$^{-1}$ both in QUMOND with $\gamma = 1$ and in the Newtonian gravity models that we investigated here. Hypervelocity stars with ejection velocities higher than this value reach galactocentric distances larger than $\sim 15$~kpc if they live long enough. We thus expect that the $|v_\phi|$ component of $4~M_{\sun}$ HVSs beyond this distance can be used to probe the deviation from the axial symmetry of the Galactic potential. 

The ejection velocity threshold and the corresponding minimum galactocentric distance increase with decreasing HVS mass. For example, in Newtonian gravity, $3~M_\sun$ and $2.5~M_\sun$ HVSs have ejection velocity thresholds of $\sim 790$ km~s$^{-1}$ and $\sim 815$ km~s$^{-1}$, which correspond to the minimum galactocentric distances of 30 kpc and 50 kpc, respectively. In QUMOND with $\gamma = 1$ ($\gamma = 2$), $3~M_\sun$ and $2.5~M_\sun$ HVSs have ejection velocity thresholds of $\sim 800$ ($755$) km~s$^{-1}$ and $\sim 830$ ($785$) km~s$^{-1}$, which correspond to the minimum galactocentric distances of 30 ($30$) kpc and 50 ($50$) kpc, respectively (see Sect.~\ref{sec:sample}). Therefore, the mass of the star sets the minimum distances beyond which the star must be located if we wish to use it for the test we suggested here. 

In Newtonian gravity, the symmetric features of the baryonic distribution can be overcome by those of the dark matter halo surrounding the Galaxy. Here, we explored the case where both the baryonic components and the dark matter halo are axisymmetric but their axes are misaligned; the shape of the halo depends on the distance from the Galactic center, and the halo approaches the spherical symmetry at distances larger than $\sim 21.6$~kpc. It follows that $v_\phi$ is affected by the non-axisymmetric features of the potential up to this radius.  

Among the baryonic components, only a triaxial bulge  as well as the possible presence of a nonspherical HG halo can affect $v_\phi$. Additional variations in $v_\phi$ might derive from massive external objects, such as the LMC \citep{kenyon2018}, whereas other non-axisymmetric components within the disk, such as spiral arms or density inhomogeneities, are expected to affect only the stars within $\sim 1$~kpc of the disk \citep{Gardner:2020jsf}. The contribution from the HG halo in $|v_\phi|$ strongly depends on its shape and total mass, which are poorly constrained by observations.  

In our QUMOND model with a triaxial bulge, the triaxiality of the bulge is effective up to $r \sim 5$ kpc; beyond this radius, the bulge can be approximated to be spherical and, because of the conservation of angular momentum, the maximum value of $\vert v_\phi \vert$ is proportional to $r^{-1}$ (Eq.~(\ref{eq:QUMONDupperlimit})). Therefore, when compared with the Newtonian model affecting $|v_\phi|$ out to $\sim 21.6$~kpc, the values of $|v_\phi|$ of HVSs at $r\gtrsim 15$~kpc in QUMOND are substantially lower than the values allowed for a non-axisymmetric potential in Newtonian gravity. For example, at $r \sim 20$ kpc, we find $|v_\phi| \lesssim$ 2 km~s$^{-1}$ in QUMOND, whereas $|v_\phi|$ can be as high as 6 or 10 km~s$^{-1}$ for a dark matter halo with axial ratios $q_y = 1.1$ or 1.2, respectively (see Fig.~\ref{fig:rvphi}). 

If a nonspherical HG halo is included in the MW model, it dominates the $|v_\phi|$ values of the HVSs in QUMOND, and the $|v_\phi|$ values in this case  may be higher than those generated  by the triaxial bulge alone. On the contrary, in Newtonian gravity, the $|v_\phi|$ values  are still dominated by the dark matter halo and can still largely exceed the QUMOND values for HVSs at distances of up to 60 kpc from the center.

We conclude that precise measurements of $v_\phi$ for a few $4~M_{\sun}$ HVSs at galactocentric distances larger than $\sim 15$~kpc (or $3~M_\sun$ HVSs at $r \gtrsim 30$ kpc, or $2.5~M_\sun$ HVSs at $r \gtrsim 50$ kpc) and smaller than $\sim 60$ kpc may in principle test the validity of QUMOND. Finding a few HVSs with azimuthal components, $|v_\phi|$, above the QUMOND upper limits, $\vert v_\phi^{\rm max}\vert$, given by Eq.~(\ref{eq:QUMONDupperlimit}) (black line in Fig.~\ref{fig:rvphi}) or shown in  Figs.~\ref{fig:rvphiHG10}-\ref{fig:rvphiHG01} in the presence of the HG halo, would suggest that MOND may not be the correct theory of gravity, at least in its QUMOND formulation. Such tests clearly require that the HVSs are confirmed to originate from the Galactic center and that their trajectories are not perturbed by external objects, such as the LMC. 

Assessing the HVS nature of observed stars is far from trivial, however, because we need to trace their trajectories back in time with the theory of gravity and in the gravitational potential that we wish to constrain. In addition, in QUMOND the perturbation of the trajectories by objects beyond the MW is complicated by the external field effect that is absent in Newtonian gravity \citep[e.g.,][]{2016MNRAS.458.4172H, 2018MNRAS.480..473F, 2020A&A...640A..26H}. The external field effect is expected to be more relevant for the most distant stars and needs to be quantified. We plan to tackle these issues elsewhere.

If we assume that the stars currently identified as HVSs are indeed HVSs in both QUMOND and Newtonian gravity, we conclude that the current uncertainties on their azimuthal velocities, $v_\phi$, mostly due to the large relative uncertainties on the proper motion measurements, are too large to provide a conclusive comparison of the data with our QUMOND limit. Indeed, in the subsample of nine HVSs drawn from the sample of \cite{brown2014} that we considered here, the azimuthal velocities have relative uncertainties in the range $35$-$340\% $. The precision on the estimates of $v_\phi$ thus needs to be improved by at least a factor  of $\sim 10$ to make our test decisive. Future measurements from space-borne astrometric missions with expected microarcsecond precision on star positions, such as Theia \citep{theia2017, malbet2019, malbet2021},  are expected to allow us to discriminate between the two theories of gravity.

\begin{acknowledgements}

We are sincerely grateful to the referee who provided a number of insightful and constructive suggestions that helped to clarify parts of our work that were initially vague. We thank Warren Brown, Margaret Geller and Scott Kenyon for stimulating discussions on the topic of hypervelocity stars, and for constructive comments on the manuscript. SSC was supported by the grant ``The Milky Way and Dwarf Weights with Space Scales'' funded by University of Torino and Compagnia di S.~Paolo (UniTO-CSP), by the grant no.~IDROL~70541 IDRF~2020.0756 funded by Fondazione CRT, by INFN, and by the Departments of Excellence grant L.232/2016 of the Italian Ministry of Education, University and Research (MIUR). 
This last grant fully supported the PhD fellowship of AG. We acknowledge partial support from the INFN grant InDark. The work of SE included here was part of his Master Thesis project at the University of Torino.
This work has made use of data from the European Space Agency (ESA) mission {\it Gaia} (\url{https://www.cosmos.esa.int/gaia}), processed by the {\it Gaia} Data Processing and Analysis Consortium (DPAC, \url{https://www.cosmos.esa.int/web/gaia/dpac/consortium}). Funding for the DPAC has been provided by national institutions, in particular the institutions participating in the {\it Gaia} Multilateral Agreement.
This research has made use of NASA’s Astrophysics Data System Bibliographic Services. This work has made use of \textsc{Topcat} \citep{Topcat2005} and the following \textsc{Python} modules: \textsc{Matplotlib} \citep{Matplotlib}, \textsc{NumPy} \citep{NumPy}, \textsc{SciPy} \citep{SciPy}, \textsc{AstroPy} \citep{astropy:2013, astropy:2018} \textsc{GalPy} \citep{2015ApJS..216...29B}.

\end{acknowledgements}

\bibliographystyle{aa}
\bibliography{hvs.bib}

\end{document}